\documentclass[aps,prb,10pt,twocolumn,eqsecnum]{revtex4}
\usepackage{graphicx,amsmath,amssymb,bm,bbm,latexsym}

\usepackage{color}
\allowdisplaybreaks

\newcommand{\be}{ \begin{equation}}
\newcommand{\ee}{ \end{equation}}

\usepackage{color}

\newcommand{\SU}[1]{\textcolor{black}{#1}}

\newcommand{\GMFP}[1]{\textcolor{black}{#1}}

\usepackage{ulem}

\begin{document}

\title{
Emergence of Dirac Electron Pair in Charge Ordered State of Organic Conductor $\alpha$-(BEDT-TTF)$_2$I$_3$}
\author{A. Kobayashi$^1$, Y. Suzumura$^1$,  F. Pi\'echon$^2$ and G. Montambaux$^2$
}

\affiliation{
$^1$Department of Physics, Nagoya University, Furo-cho, Chikusa-ku, Nagoya, 464-8602 Japan \\
$^2$Laboratoire de Physique des Solides, CNRS UMR 8502, Universite Paris-Sud, F-91405 Orsay Cedex, France
}

\date{\today}

\begin{abstract}
 We  re-examine the band structure of the stripe charge ordered state of $\alpha$-(BEDT-TTF)$_2$I$_3$
 under pressure by using an extended Hubbard model within the  Hartree mean-field theory.
  By increasing pressure, we find a topological transition from a conventional insulator
with a  single-minimum in the dispersion relation at the M-point in the Brillouin zone, towards a new phase
which exhibits a double-minimum. This transition is characterized  by  the appearance of a pair  of
Dirac electrons with a finite mass.  Using the Luttinger-Kohn representation at the M-point,
it is shown that such a variation of the band structure can be
described by an effective $2 \times 2$  low energy Hamiltonian with a single  driving parameter.  The topological nature of this transition is confirmed by the calculation of the
Berry curvature which vanishes in the conventional phase and has a double peak structure with opposite signs in the new phase. We compare the structure of this transition with a simpler situation which occurs in two-component systems, like boron-nitride.
\end{abstract}


\maketitle

\section{Introduction}

Two-dimensional molecular conductor $\alpha$-(BEDT-TTF)$_2$I$_3$
\cite{TajimaRev2009} has brought
 much interest by the variety of electronic states, such as an insulating stripe  Charge Ordered (CO) state
\cite{SeoRev2004}, a superconducting state in the presence of charge ordering, and a Zero Gap State (ZGS)
 with a massless Dirac spectrum \cite{KobayashiRev2009}.

The stripe CO state \cite{Kino-Fukuyama,Seo,Hotta}, which was suggested  to explain an insulating phase below $135$K at ambient pressure, was confirmed by NMR experiment \cite{TakahashiStripe}.
The superconducting state, found under uniaxial pressure along stacking axis ($a$-axis) \cite{Tajima-Ebina2002}, was investigated theoretically by using an extended Hubbard model \cite{Kobayashi2005}.
A narrow gap state was  suggested by Kajita to explain anomalous increase of Hall coefficient at high pressures \cite{KajitaFirst}.
The existence of massless Dirac electrons  was predicted theoretically \cite{Katayama2006ZGS} based on a tight-binding calculation using the transfer energies
of ref. \onlinecite{Kondo2005}.
Thus it has been revealed that this narrow gap state is indeed a zero gap state, as  also confirmed by a first principle calculation \cite{Kino,Ishibashi}.
 In this state, the energy spectrum near the Fermi energy consists of two cones described by a tilted Weyl equation for massless Dirac electrons \cite{Kobayashi2007,Montambaux2008TiltedWeyl}.
 This has been
confirmed by a comparison between the theoretical and experimental
results for the temperature dependence of
the Hall coefficient \cite{Kobayashi2008,Tajima2009}
and the angular dependence of the magnetoresistance \cite{Tajima2009,Morinari2009}.

 The two tilted Dirac cones were predicted to merge and disappear at the $\Gamma$ point under extremely high pressure \cite{Kobayashi2007}. This merging transition was also studied in the context of deformed graphene with a  variation of transfer integrals, and was described
 using a generalized two-component Hamiltonian for Dirac electrons.
 This effective Hamiltonian describes the merging of two Dirac cones
 and the opening of a gap at the transition \cite{Dietl,Montambaux2009EPJ,Montambaux2009PRB}.

 In the present paper, we show that  another type of transition may also occur in the CO state of $\alpha$-(BEDT-TTF)$_2$I$_3$.
 In this state, the energy spectrum exhibits a  gap between the conduction and
 valence bands at the M-point, with a single minimum. By increasing pressure, it is demonstrated that  new electronic phases
emerge  in the CO state with a double-minimum structure in the vicinity of
the M-point.  This double-minimum corresponds to the emergence of a pair
of massive Dirac electrons,  whose study is the main goal of this paper.
This qualitative change of the band structure is described by an
effective Hamiltonian with a single driving parameter.

The paper is organized as follows.
In section 2, a tight binding model for $\alpha$-(BEDT-TTF)$_2$I$_3$
 is described  where the repulsive interactions between molecules are treated within Hartree mean-field theory.
 When varying  pressure and the intersite
repulsive interaction, a new phase diagram is obtained, and is
described  in section 3, where
new phases characterized by a Dirac electron pair are obtained in the CO state.  In section 4, using the Luttinger-Kohn representation at the M-point of the Brillouin zone, we construct a $2 \times 2$ effective Hamiltonian to describe the low energy band structure near this point, and the emergence of a pair of Dirac points.  The general structure of this effective Hamiltonian also describes the emergence of Dirac points in a simple toy-model related to the physics of boron nitride (BN).
 In section 5, we show that  this Dirac pair  is revealed by
two sharp peaks of opposite sign in the Berry curvature
\cite{Berry}. Then each Dirac point is characterized by an appropriate Berry phase  calculated  by using a method already  applied to
  the physics of  Dirac points in boron nitride \cite{Fuchs2010}.
The last section is devoted to summary and discussion.

\section{Tight-binding model for $\alpha$-(BEDT-TTF)$_2$I$_3$ }

\subsection{The Hamiltonian}

\begin{figure}
\includegraphics[height=60mm]{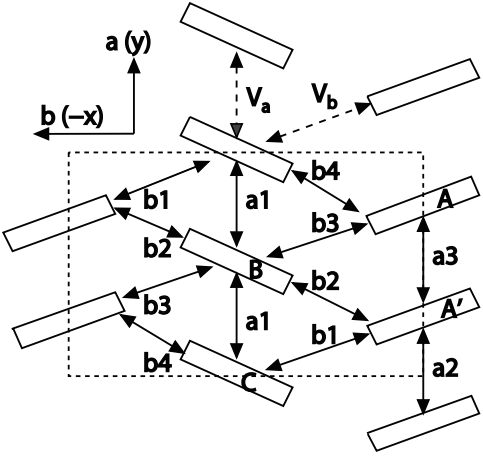}
\caption{\label{unitcell}
The model describing the electronic system of $\alpha$-(BEDT-TTF)$_2$I$_3$ \cite{Kino,Mori1984,Mori1999}.
The unit cell consists of four BEDT-TTF molecules ${\rm A}$, ${\rm A}^\prime$, ${\rm B}$ and ${\rm C}$ with  seven   transfer energies.
The nearest neighbor repulsive interactions are given by $V_a$ and $V_b$.
The $a$- and $b$-axis in the conventional notation correspond to the $y$- and $x$-axis in the present paper.
}
\end{figure}

The model  used to describe the two-dimensional electronic system in
$\alpha$-(BEDT-TTF)$_2$I$_3$ is shown in
Fig.~\ref{unitcell} \cite{Kino,Mori1984,Mori1999}. The unit cell
consists of four BEDT-TTF molecules
   on sites A, ${\rm A}^\prime$, B and C.
The sites A, B and C are inequivalent,
   while  A is equivalent to  A' so that inversion symmetry is preserved.
There are six electrons for the four molecules in a unit cell,
{\it i.e.}, the  band  is  $3/4$-filled. On the basis of the HOMO orbitals of
these sites \cite{Kino-Fukuyama,Seo}, these  electrons are described
   by the extended Hubbard model with several transfer energies,
    the on-site repulsive interaction $U$
    and the anisotropic nearest-neighbor
       repulsive interaction $V_{\alpha \beta}$~:
\begin{eqnarray}
H &=& \sum_{( i \alpha : j \beta ),
\sigma}
 (t_{i \alpha; j \beta}\
   a^{\dag}_{i\alpha\sigma}a_{j\beta\sigma}+ {\rm h. c.} )
     \nonumber\\
 &+& \sum_{i\alpha}
    U\ a^{\dag}_{i\alpha\uparrow}a^{\dag}_{i\alpha\downarrow}
      a_{i\alpha\downarrow}a_{i\alpha\uparrow}
     \nonumber\\
  &+&     \sum_{( i\alpha:j\beta ), \sigma, \sigma^\prime}
  V_{\alpha\beta} \ a^{\dag}_{i\alpha\sigma}a^{\dag}_{j\beta\sigma^\prime}
a_{j\beta\sigma^\prime}a_{i\alpha\sigma} ,
\label{h1}
\end{eqnarray}
where $i, j$ denote site indices of a given unit cell,
   and $\alpha, \beta (={\rm A}$, ${\rm A}^\prime$, ${\rm B}$ and ${\rm C}$) are indices of BEDT-TTF sites
    in the unit cell.
Hereafter, the energies are given in eV.
In the first term,
   $a^{\dag}_{i\alpha\sigma}$ denotes a creation operator
    with spin $\sigma (=\uparrow ,\downarrow)$ and
    $t_{i\alpha ; j \beta}$ is the transfer
       energy between the $(i,\alpha)$ site and the  $(j,\beta)$ site.

The transfer energies as a function of
  a uniaxial pressure ($P_a$) along the $a$-axis
     are estimated from  an extrapolation formula given by
\begin{equation}
t_X (P_a) = t_X (0) ( 1 + K_X P_a) .
\end{equation}
The transfer energies $t_X$ and the coefficients $K_X$
   are obtained from the numerical data
   at $P_a = 0$kbar \cite{Mori1984,Mori1999}
      and at $P_a =2$kbar \cite{Kondo2005}, such that
$
t_{a1}(0) = -0.028,
t_{a2}(0) = 0.048,
t_{a3}(0) = -0.020,
t_{b1}(0) = 0.123,
t_{b2}(0) = 0.140,
t_{b3}(0) = -0.062,
t_{b4}(0) = -0.025,
$ [eV] and
$
K_{a1} = 0.089,
K_{a2} = 0.167,
K_{a3} = -0.025,
K_{b1} = 0,
K_{b2} = 0.011,
K_{b3} = 0.032,
K_{b4} = 0
$ [eV/kbar].
We use the parameter $U=0.4$  and the $V_{\alpha\beta}$ take two different values, $V_a =0.17 \sim 0.18$ eV along the stacking direction, and $V_b=0.05$ eV along the perpendicular direction
\cite{KobayashiRev2009}. With this choice of parameters, we obtain a pressure dependence of the electronic spectrum
      consistent with experimental
      results \cite{Kobayashi2007,Tajima-Ebina2002}. Moreover
  the results of the mean-field theory
    coincide with the experimental results
       for  the charge disproportionation
           in the ZGS and the charge ordering in the CO phase
\cite{Kakiuchi2007,Takahashi2008CD}.
  By "charge disproportionation", we mean that
   the A, B and C sites in a unit cell are inequivalent but
     the inversion symmetry between A and A' remains.  By "charge ordered" state, we mean that the inversion symmetry
 between A and A' is broken. Throughout the paper, $\hbar$ and the lattice constant $a$ are taken as unity.

\subsection{Hartree mean-field theory}

 As in previous works \cite{Kobayashi2004_11,Kobayashi2005}, we restrict ourselves to a Hartree mean field theory which implies that the mean field Hamiltonian is  diagonal in spin space. In addition we only consider mean field solutions that do not break the underlying Bravais lattice symmetry; this further implies that for a given spin $\sigma$ the mean field Hamiltonian $H_{\sigma}({\bf k})$ is a $4\times4$ matrix in the Bloch-basis $a_{{\bf k}\beta\sigma}$. In second quantized form, the mean-field Hamiltonian reads :
\begin{eqnarray}
H_{MF}&=&\int_{BZ}{ d{\bf k} \over (2 \pi)^2} \sum_{\sigma} H_{\sigma}({\bf k}),\\
H_{\sigma}({\bf k})  &=& \sum_{\alpha\beta}
\tilde{\epsilon}_{\alpha \beta \sigma}({\bf k})
 a^{\dag}_{{\bf k}\alpha\sigma}a_{{\bf k}\beta\sigma},\\
   \label{eq:MF_Hamiltonian}
    \nonumber   \\
 \tilde{\epsilon}_{\alpha \beta \sigma}({\bf k})
   &=& I_{\alpha \sigma}  \delta_{\alpha \beta}
         + \epsilon_{\alpha \beta}({\bf k}),
\label{hmf}
      \\
  I_{\alpha \sigma}  &=&
U_\alpha \langle n_{\alpha -\sigma} \rangle
     +
    \sum_{\beta' \sigma'}
     V_{\alpha \beta'} \langle n_{\beta' \sigma'} \rangle,   \\
\epsilon_{\alpha\beta}({\bf k})
&=& \sum_{{\bf \delta}}
t_{\alpha\beta}
e^{{\rm i} {\bf k}\cdot { \delta }},
\end{eqnarray}
where $\langle n_{\alpha \sigma} \rangle = \langle a_{i \alpha \sigma}^\dagger a_{i \alpha \sigma}\rangle$
is the mean field local density of spin $\sigma$   for the molecular state $\alpha$ in unit cell $i$;
$\langle n_{\alpha \sigma} \rangle$ is  assumed to be independent of the unit cell $i$.
${\boldmath \delta}$ denotes the vectors connecting nearest neighbors sites.
The spin dependent site potential $I_{\alpha \sigma}$ represents the Hartree mean field
  which comes from the on-site Hubbard $U$ and nearest-neighbor Coulomb interactions $V_a,V_b$.

Note that off-diagonal elements $\epsilon_{\alpha \beta}({\bf k})$ are independent on spin index $\sigma$ and on interaction strength but depend on ${\bf k}$ and $P_a$. Moreover these off-diagonal elements $\epsilon_{\alpha \beta}({\bf k})$ have the time reversal symmetry
$\epsilon_{\alpha \beta}^*({\bf k})=\epsilon_{\alpha \beta}(-{\bf k})$.
In this representation, only the self-consistent diagonal elements may explicitly exhibit the breaking of time reversal and/or inversion symmetries.

To obtain the mean-field phase diagram,  the Hamiltonian is diagonalized numerically for a given ${\bf k}$ in each spin subspace, according to

\begin{eqnarray}
\sum_{\beta=1}^{4}
\tilde{\epsilon}_{\alpha\beta\sigma}({\bf k})\ d_{\beta \gamma \sigma}({\bf k})
&=&\xi_{\gamma \sigma}({\bf k})\ d_{\alpha \gamma \sigma}({\bf k}) ,
\label{eigenvalue}
\end{eqnarray}
where $\xi_{\gamma \sigma}$ are the eigenenergies ordered such that,
$
 \xi_{1 \sigma}({\bf k}) > \xi_{2 \sigma}({\bf k})
  > \xi_{3 \sigma}({\bf k}) > \xi_{4 \sigma}({\bf k})
$  ($\gamma=1,2,3,4$  is the  band index), and $d_{\alpha \gamma \sigma}({\bf k})$  are
the corresponding eigenvectors.
In terms of eq.~(\ref{eigenvalue}),
the average number $\langle n_{\alpha\sigma} \rangle$ of electrons  with spin $\sigma$ on $\alpha$ type of site
is expressed as
\begin{eqnarray}
\langle n_{\alpha \sigma} \rangle &=&\int_{BZ} {d{\bf k} \over (2 \pi)^2} \sum_{\gamma =1}^{4}
d^{*}_{\alpha \gamma \sigma}({\bf k})\ d_{\alpha \gamma \sigma}({\bf k})
n_{F}(\xi_{\gamma \sigma}({\bf k})),
         \nonumber \\
\label{sceq}
\end{eqnarray}
where $n_{F}(\xi_{\gamma \sigma}({\bf k}))=
1/(\exp{[(\xi_{\gamma \sigma}({\bf k})-\mu)/T]}+1)$  is the Fermi factor at temperature $T$ ($k_{B}=1$) and $\mu$ is the chemical potential
determined  from the condition of a 3/4 filled system:
\begin{eqnarray}
\frac{3}{4}&=&\frac{1}{8} \sum_{\alpha\sigma} \langle n_{\alpha\sigma} \rangle=\frac{1}{8} \int_{BZ} {d{\bf k} \over (2 \pi)^2} \sum_{\gamma \sigma} n_{F}(\xi_{\gamma \sigma}({\bf k})).
 \nonumber \\
\end{eqnarray}
Equation ~(\ref{sceq}) constitutes the self-consistent relation for the mean field quantities
 $\langle n_{\alpha\sigma} \rangle$.
A high accuracy in fulfilling these constraints requires a very fine mesh of the reciprocal space which constitutes the main numerical difficulty.
When convergence is achieved, the mean field energy per unit cell of a given state is then calculated as
\begin{eqnarray}
E_{\rm MF}&=&\int_{BZ} {d{\bf k} \over (2 \pi)^2} \sum_{\gamma} n_{F}(\xi_{\gamma \sigma}({\bf k})) \xi_{\gamma \sigma}({\bf k})
 \nonumber \\
 &-&
 \sum_{\alpha} U_{\alpha}
   \langle n_{\alpha \uparrow} \rangle \langle n_{\alpha \downarrow} \rangle
  - \sum_{( \alpha, \beta ), \sigma, \sigma'}
     V_{\alpha \beta}
      \langle n_{\alpha \sigma} \rangle
       \langle n_{\beta \sigma'} \rangle.
 \nonumber \\
\end{eqnarray}
 The ground state is obtained
 from minimization of $E_{\rm MF}$.

\section{Phase diagram}


\begin{figure}
\includegraphics[height=80mm]{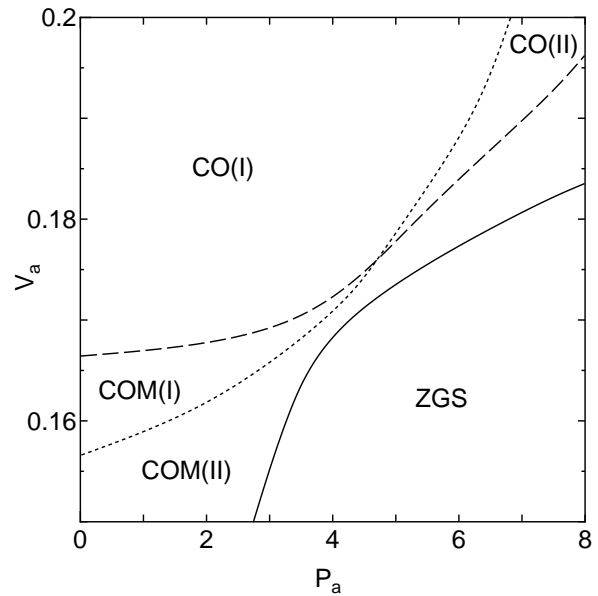}
\caption{\label{PhaseDiagram}
The phase diagram on the uniaxial pressure along $a$-axis ($P_a$ [kbar]) and the repulsive interaction between nearest-neighbor sites along $a$-axis ($V_a$ [eV])
where $U$ = 0.4 eV, and $V_b$ = 0.05 eV.
 The CO and COM denote insulating and metallic states respectively.
In addition to the phase (I) of the previous work\cite{Kobayashi2005},
 there exists a new phase  (II), which  is characterized by a double minimum in the up spin band.
 }
\end{figure}
\begin{figure}
\includegraphics[height=80mm]{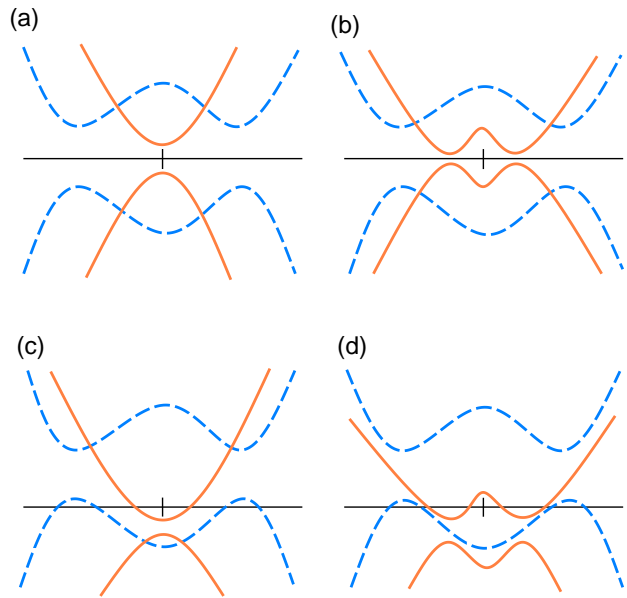}
\caption{\label{DiagramBand}
Schematic behavior of the energy spectrum close to
the Fermi energy (horizontal line), in the different subphases of the CO phase : (a)CO(I), (b)CO(II), (c)COM(I), (d) COM(II). The red   and blue bands correspond respectively to $\uparrow$ and $\downarrow$ spins.
The center of the horizontal line corresponds to the  M point (${\bf k}_M= (\pi, \pm \pi)$).}
\end{figure}

The ($P_a$-$V_a$) phase diagram obtained from the self-consistent Hartree approximation described above is presented in Fig.~\ref{PhaseDiagram}.  $P_a$ is the uniaxial pressure along the $a$-axis and $V_a$ is the repulsive interaction
between nearest-neighbor sites along the $a$-axis. This figure exhibits three transition lines. Two of them (the continuous and dashed lines) were already found in  previous works \cite{Kobayashi2005} and the third one (dotted line) is a novel transition that constitutes the main object of this work.
The schematic band spectrum close to the Fermi surface is shown
in Fig.~\ref{DiagramBand}
and is discussed below.
 Before describing this last transition in more details, we remind the qualitative picture associated to the two other transitions.

In Fig.~\ref{PhaseDiagram}, the solid line  marks a charge ordering transition resulting from the simultaneous breaking of time reversal and inversion symmetries. It separates  the charge ordered metallic state (COM) from the Zero Gap state (ZGS). In the ZGS phase, energy bands are spin degenerate and inversion symmetry is preserved such that conduction band $\xi_{1 \sigma}({\bf k})$ and valence band $\xi_{2 \sigma}({\bf k})$ touch each other at two Fermi points (Dirac points) ${\bf k}_\pm$. These Dirac points move in the Brillouin zone when varying parameters $P_a$ and $V_a$. Around each point $ {\bf k}_\pm$, the dispersion relation is linear. In addition  there is a large anisotropy in the Fermi velocities (a factor $\sim 10$ between the highest and lowest velocity values  \cite{Katayama2006ZGS,Kino}), this  appears as a tilt in the Dirac cones \cite{Kobayashi2007}.
Coming from this high pressure ZGS phase and traversing the continuous line, the inversion symmetry is spontaneously broken by the electronic interactions. As consequence, for a given spin $\sigma$, a gap opens between bands $\xi_{1 \sigma}({\bf k})$ and $\xi_{2 \sigma}({\bf k})$, leading {\it a priori} to an insulating phase.
However the time reversal symmetry is also spontaneously broken by the interactions so  that the degeneracy between $\uparrow$ and $\downarrow$ bands is now lifted. Therefore the {\it simultaneous} breaking of time reversal and inversion symmetries results in a semi-metallic phase (COM) with band overlap leading to small electrons and holes pockets of opposite spin orientations (Fig.~\ref{DiagramBand}).

In striking contrast with the continuous line,
the dashed line marks a  metal-insulator transition from a charge ordered metallic (COM) phase to a charge ordered insulator (CO) without breaking of any symmetry.
In traversing this transition line, the dispersion relations of the four energy bands $\xi_{1 \sigma}({\bf k})$ and $\xi_{2 \sigma}({\bf k})$ stay similar but their relative positions to the Fermi level vary in  such a way that, in the CO phase, the Fermi level falls in a true charge gap that separates a valence band and a conduction band of equal spin orientation (Fig.~\ref{DiagramBand}).
We stress that, at the metal-insulator transition, the electron density $\langle n_{\alpha \sigma} \rangle$ and therefore the resulting mean field potential $I_{\alpha\sigma}$ exhibit a cusp
  on respective sites.

\begin{figure}[htbp]
 \begin{minipage}{0.5\hsize}
  \begin{center}\leavevmode
   \hspace*{0cm}\includegraphics[width=5cm]{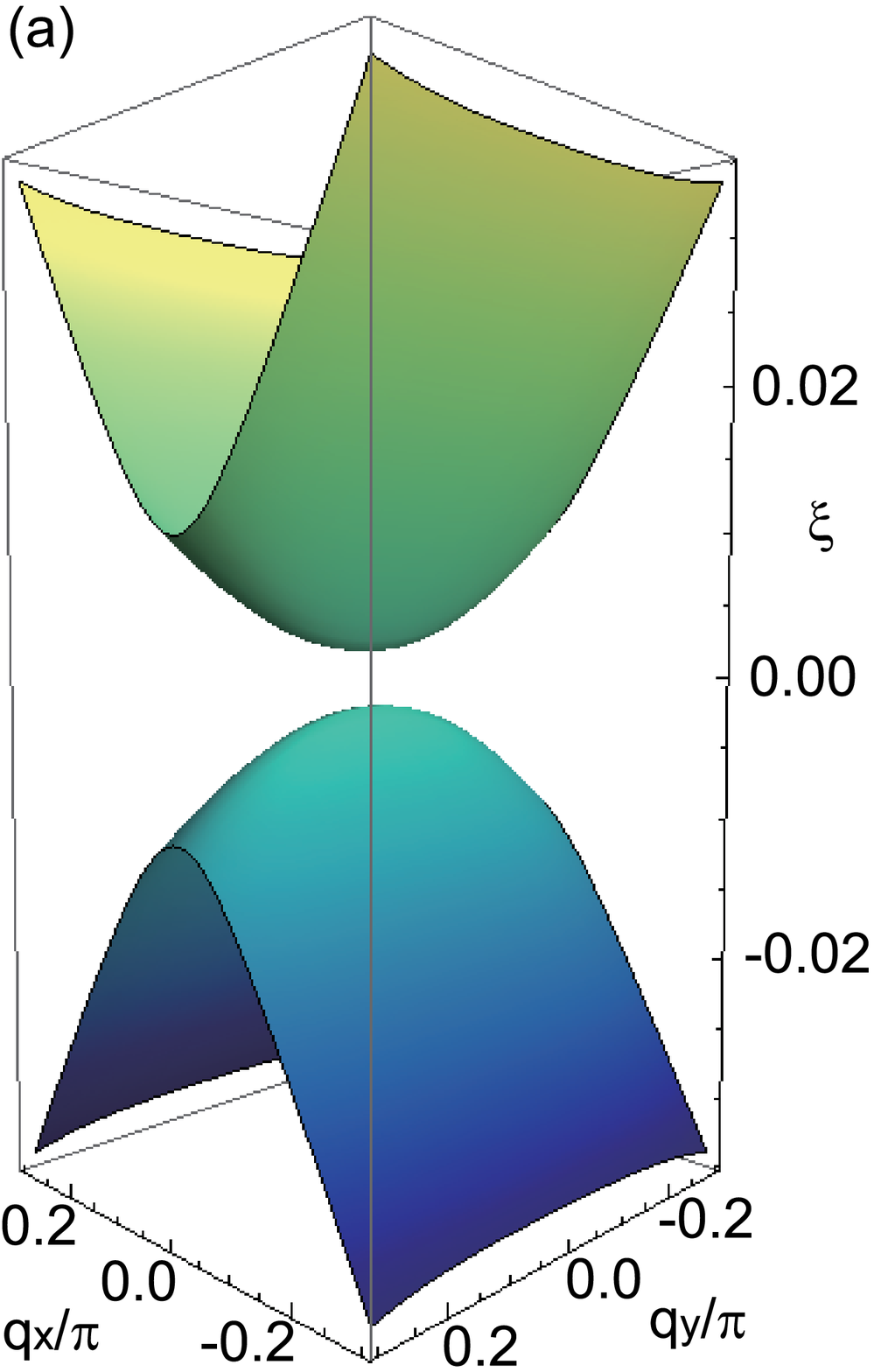}
    \end{center}
   \end{minipage}
 \begin{minipage}{0.5\hsize}
  \begin{center}\leavevmode
    \hspace*{0cm}\includegraphics[width=5cm]{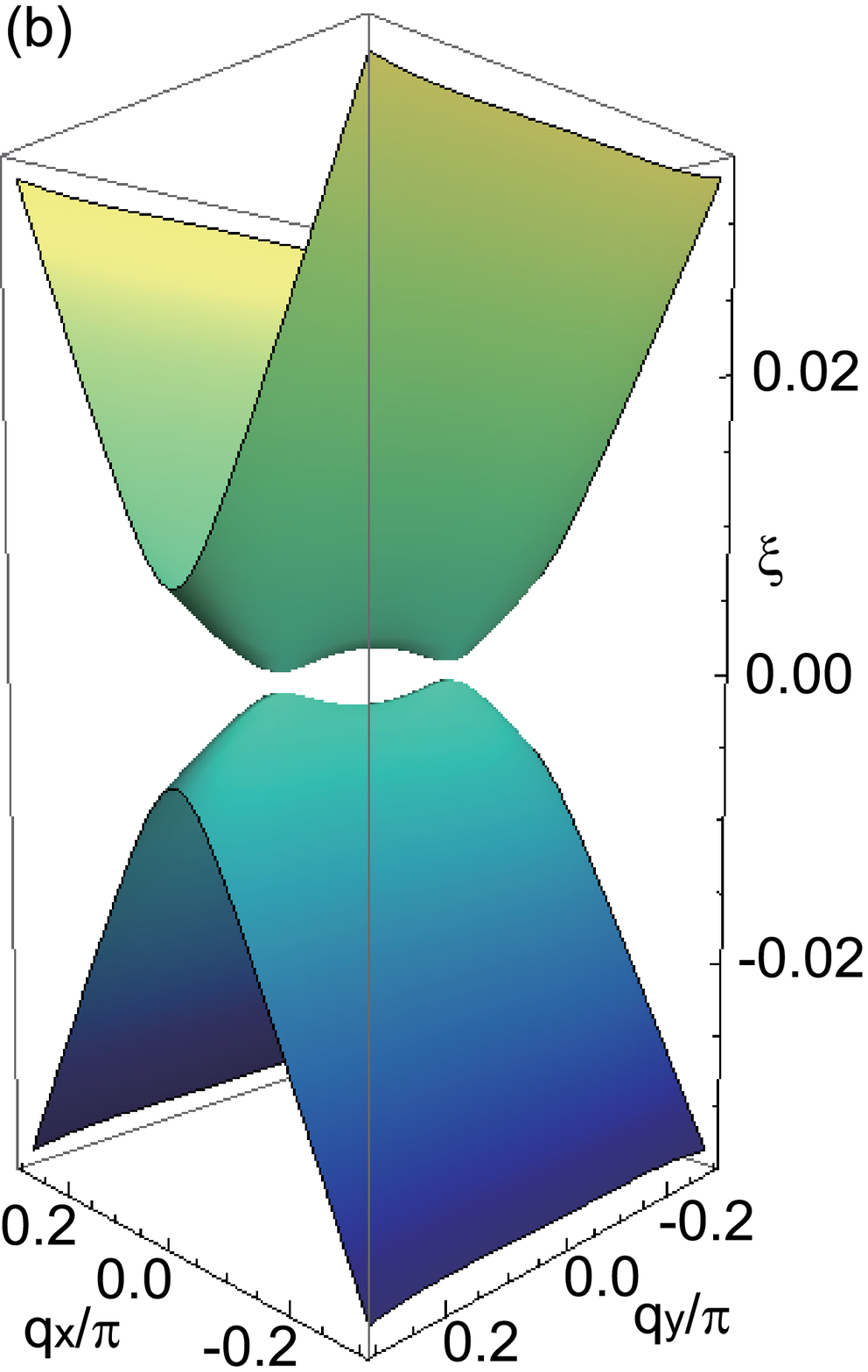}
   \end{center}
  \end{minipage}
 \caption{
The ${\bf q}$-dependence of the conduction and valence bands for
$\uparrow$ spin where ${\bf q} = {\bf k} - {\bf k}_M$, and the center denotes the M point (${\bf k}_M = (\pi, \pm \pi)$).
\GMFP{
 a) Spectrum of the CO(I) phase with a single minimum   at the  M point. b)  Spectrum of the CO(II) phase with a double minimum   around the  M point.
}
Here the parameters are $V_a =0.18$ eV and $V_b =0.05$ eV, and the pressures are  respectively  $P_a = 4.5$ kbar and $P_a =5.4$ kbar, for the  CO(I) and CO(II) phases.}
 \label{EPM}
\end{figure}

In this work, by a more detailed analysis of the COM and CO phases, we find a new {\it topological} transition (dotted line  in Fig. \ref{PhaseDiagram}) that further splits each of the COM and CO phase into two phases: COM(I,II) and CO(I,II). The electronic spectrum in each phase is represented schematically in Fig.~\ref{DiagramBand} .
 The  band structure
  of both $\uparrow$ band  and $\downarrow$ band  in CO (II) state
 is  explained in Appendix A.
This transition concerns a modification in the two energy bands close to the Fermi energy. They correspond to  a given value of the spin, that we choose to denote by
 $\uparrow$. The two other bands  ($\downarrow$) are not concerned by this transition.
As illustrated in Fig.~\ref{EPM}, the transition from CO(I) (Fig.~\ref{EPM} a) to CO(II) (Fig.~\ref{EPM} b) is characterized by a change in the form of the dispersion relation of the valence and conduction bands. In the CO(I) phase, there is a single minimum of charge gap whose position in ${\bf k}$ space stays at the M-point ${\bf k}_M=(\pi,\pm \pi)$,
independently of the parameters $P_a,V_a$. Around this point ${\bf k}_M$, valence and conduction bands disperse quadratically. In the CO(II) phase, the single charge gap separates in two points at symmetrical positions (For example, ${\bf k}_{\pm} =\pm (0.95 , -0.71 ) \pi $ for $P_a=5.4$ kbar and $V_a=0.18$ eV) from the M-point, and ${\bf k}_{\pm}$  move continuously with parameters $P_a,V_a$. There is now a double-well structure in the dispersion relation.
The aim of this work is to describe this  topological transition in the framework of a  universal Luttinger-Kohn Hamiltonian.

As a last remark concerning the phase diagram of Fig.~\ref{PhaseDiagram}, we note
 an intersection of the dashed line and the dotted line, at the critical point of coordinates
$(P_a^*, V_a^*)$  =(4.75 kbar, 0.175 eV).
 Such crossing implies the proximity of   several minima of the total   energy in its neighborhood. Therefore   in this region of the phase diagram one may need to go beyond Hartree mean-field approximation. This might be an important issue since this critical point $P_a ^*,V_a ^*$ is in a region of parameters value that is reachable experimentally.

\section{Topological transition  around the M-point}

\subsection{Low energy $2\times 2$ effective Hamiltonian at the M-point}

In order to analyze the band structure in the CO state,
we expand the mean-field Hamiltonian up to the  second order
in momentum around the M-point, writing ${\bf k}={\bf k}_M + {\bf q}$.   The Hamiltonian is diagonal in spin space and it has the $4 \times 4$ structure, for each spin direction $\sigma$~:
\begin{eqnarray}
H_{\alpha \beta \sigma} ({\bf k}) & \cong& H_{\alpha \beta \sigma} ({\bf k}_M )
  \nonumber \\
& &+ q^\mu \partial_\mu H_{\alpha \beta \sigma} ({\bf k}) \vert_{{\bf k}_M}
 +  {1 \over 2} q^\mu  q^\nu \partial_\mu \partial_\nu  H_{\alpha \beta \sigma} ({\bf k} ) \vert_{{\bf k}_M}  \nonumber \\
  = & & H_{\alpha \beta \sigma} ({\bf k}_M )
 + \delta H_{\alpha \beta \sigma} ({\bf q}).
\end{eqnarray}
We wish to restrict ourselves to the two  bands which are close to the Fermi level ($\sigma= \uparrow$). This can be done by using the Luttinger-Kohn representation \cite{LuttingerKohn} at the M-point, and we
 obtain an effective $2 \times 2$  Hamiltonian with the matrix elements
($\nu, \nu'$ = 1,2)  up to O($q^2$)~:
\cite{Katayama_EPJBS}
\begin{eqnarray}
&& H_{\nu, \nu'} = \xi^0_\nu \delta_{\nu \nu'}+ \langle \nu| \delta H ({\bf q}) | \nu'\rangle
  +  {1 \over 2}    \sum_{\nu"=3,4}
 \nonumber \\
&&
\langle \nu|\delta H ({\bf q})|\nu"\rangle
\langle  \nu"|\delta H ({\bf q})|\nu'\rangle \left( \frac{1}{\xi^0_{\nu} - \xi^0_{\nu"}}+\frac{1}{\xi^0_{\nu'} - \xi^0_{\nu"}}\right)
 \;,  \nonumber \\
\label{effectiveH22}
\end{eqnarray}
where
$\xi_\nu^0$ and $|\nu \rangle$ are respectively the eigenvalues and the eigenvectors of $H({\bf k}_M)$.
Thus the effective Hamiltonian  is rewritten as
\begin{equation}
{H}^{LK}({\bf k})={H}^{LK}({\bf k}_M)+i{\bf q} \cdot {\bf {V}} ({\bf k}_M) +\sum_{ij} {W}_{ij}({\bf k}_M) q_{i} q_{j} ,
\label{Luttinger-Kohn-H}
\end{equation}
where, by construction, ${H}^{LK}({\bf k}_M)$ is diagonal. Like the original Hamiltonian, the  Luttinger-Kohn (L-K)  Hamiltonian has the symmetry ${H^{LK}}^*({\bf k})=H^{LK}(-{\bf k})$.  Therefore, near the M-point as well as in the vicinity of the other so-called time-reversal points of position ${\bf G}/2$ where ${\bf G}$ is a reciprocal lattice vector, the energy bands $\xi_{\alpha}({\bf k})$ have peculiar properties~: for any band $\xi_{\alpha}({\bf k})$, the particular values $\xi_{\alpha}({\bf G}/2)$ are either a local extremum or a saddle point. As a consequence, the gap between two consecutive bands at these symmetry points is also either a local extremum or a saddle point.
This picture explains the topological transition from the CO(I) phase where the gap at M is a local minimum to the CO(II) phase where the gap at M is a local saddle point \cite{remark1}.  To be more precise, let us focus on the structure of the Luttinger-Kohn Hamiltonian (\ref{Luttinger-Kohn-H}).  A strong consequence of the symmetry ${H^{LK}}^*({\bf k})=H^{LK}(-{\bf k})$ is that at the  M-point, like at the other symmetry points,  $W_{ij}({\bf k}_M)$ are real symmetric matrices and  $V_{i}({\bf k}_M)$ are  real antisymmetric matrices.
We deduce  that the minimal  form of the   $2 \times 2$ matrix ${H}^{LK}({\bf k}_M+{\bf q})$
 to order $q_i q_j$ can be recast as:

\begin{equation}
{H}^{LK}({\bf k}_M+{\bf q})=f_0({\bf q})  \sigma_{0}+  f_1({\bf q}) \sigma_1 +f_2({\bf q})  \sigma_{2}+  f_3({\bf q}) \sigma_3  \; ,
\label{KLhamiltonian} \end{equation}
   where we have used a representation in terms of Pauli matrices $\sigma_i$ and

\begin{eqnarray}
\label{matrix_f0}
f_0 ({\bf q})&=&\mu +\sum_{ij}   w^{ij}_0 \ q_i q_j
 \; , \\
\label{matrix_f1}
f_1 ({\bf q})&=& \sum_{ij}  w^{ij}_1 \ q_i q_j \; , \\
\label{matrix_f2}
f_2 ({\bf q})&=&{\bf v} \cdot {\bf q}\; , \\
\label{matrix_f3}
f_3 ({\bf q})&=& \Delta+\sum_{ij}w^{ij}_3 \ q_i q_j \; .
\end{eqnarray}
with $\Delta >0$.
The eigenenergies near the  M-point are given by
\begin{equation}
\xi_{\pm}({\bf q})=f_0({\bf q}) \pm \sqrt{ f_1({\bf q})^2 + f_2({\bf q})^2+f_3({\bf q})^2} \; .
 \end{equation}
The local band structure around M has the symmetry $\xi_{\pm}({\bf q})=\xi_{\pm}({-\bf q})$. Moreover these expressions also imply that $\partial_{q_i} \xi_{\pm}({\bf q})|_{{\bf q=0}}=0$ so that, as anticipated, M is either a local extremum or a saddle point of the dispersion relation of each band.
The same properties are also valid for the gap
 $\Delta({\bf q})=\frac{1}{2}(\xi_+({\bf q})-\xi_-({\bf q}))$ separating the two bands.

\subsection{CO(I)-CO(II) transition: phase boundary}

Within the framework of this Luttinger-Kohn representation, we now determine the condition for a transition from a minimum to a saddle point at  the M-point.
For this purpose, it is useful to define the $2\times 2$ stability
(Hessian) matrix $S_{M}$. We have
\begin{equation}
\Delta({\bf q}) = \frac{\xi_+({\bf q})-\xi_-({\bf q})}{2}
\simeq \Delta + \frac{1}{2} (q_x \ q_y)S_{M}
\left(\begin{array}{l}
q_x\\
q_y
\end{array} \right) + \cdots \; ,
\label{stability1}
\end{equation}
where

\begin{equation}
S_{M}\equiv
 \left( \begin{array}{ll}
\frac{\partial^2 \Delta({\bf q})}{\partial_{q_{x}}^2} &
\frac{\partial^2 \Delta({\bf q})}{\partial_{q_{x}} \partial_{q_{y}}}\\
\frac{\partial^2 \Delta({\bf q})}{\partial_{q_{y}} \partial_{q_{x}}}&
\frac{\partial^2 \Delta({\bf q})}{\partial_{q_{y}}^2}
\end{array}\right)_{M}
=\left( \begin{array}{cc}
\overline{w}^{xx} _3& \overline{w}^{xy} _3\\
\overline{w}^{xy} _3& \overline{w}^{yy} _3
\end{array}\right) \; ,
\end{equation}
with $\overline{w}^{ij} _3=w^{ij}_3+\frac{c_{i} c_j}{2\Delta}$. The determinant of this matrix:

\begin{equation}
 \textrm{det} S_M= \overline{w}^{xx}_3 \overline{w}^{yy}_3 -  ( \overline{w}^{xy}_3)^2 \; ,
\label{eq:singleparameter}
\end{equation}
governs the stability of the M-point. There is an extremum when $\mbox{det} S_{M} > 0$ and a saddle point when
$ \textrm{det}S_M  < 0$. Therefore the transition line is given by the condition $\textrm{det}S_M = 0$, that is

\begin{equation}
\left( w_3^{xx} + { v_x^2 \over 2 \Delta} \right) \left( w_3^{yy} + { v_y^2 \over 2 \Delta} \right)= \left({w_3^{xy}+ { v_x v_y \over 2 \Delta}             }\right)^2 \; .
\label{dettrans} \end{equation}

We emphasize that the stability matrix $S_{M}$ is totally independent of the terms $w^{ij}_1$. Nevertheless, these terms are important to determine the position of the Dirac points and the size of the gap at the Dirac points in the CO(II) phase. They also play a crucial role in determining the topological properties of the band structure (Berry curvature). Therefore, among the  9 initial parameters of Eqs.~(\ref{matrix_f1}), (\ref{matrix_f2}) and (\ref{matrix_f3})
, only $6$ are pertinent to determine the transition line.
They enter  the combination $\mbox{det} S_M$ which is the {\it single driving parameter} for the CO(I) - CO(II) transition.

\subsection{Minimal form of the effective Hamiltonian for the merging transition}

In order to analyse the local structure of the Hamiltonian near a time-reversal point, it is convenient to  parametrize it using polar coordinates of the wave vector ${\bf q}$. We write~:
\begin{equation}
q_x = q \cos \theta \qquad  , \qquad q_y = q \sin\theta \; ,
\end{equation}
so that the Hamiltonian describing the CO(I)-CO(II) transition can be written in the form (since the component $\sigma_0$ plays no role in our discussion, we define the effective Hamiltonian as $h({\bf q})= H^{LK}({\bf k}_M + {\bf q}) - f_0({\bf q}) \sigma_0)$~:

\begin{equation}
h({\bf q}) = \left(
               \begin{array}{cc}
                 \Delta + w_3^\theta q^2  & - i v_\theta q + w_1^\theta q^2  \\
                 & \\
                  i v_\theta q + w_1^\theta q^2  & -\Delta - w_3^\theta q^2 \\
               \end{array}
           \right) \; ,
\label{eq:hqtheta} \end{equation}
where
\begin{equation}
\begin{array}{l}
w^{\theta} _1=w^{yy} _1 \sin^2{\theta}+w^{xx} _1\cos^2{\theta}+2w^{xy} _1\cos{\theta}\sin{\theta},\\
w^{\theta} _3=w^{yy} _3 \sin^2{\theta}+w^{xx} _3\cos^2{\theta}+2w^{xy} _3\cos{\theta}\sin{\theta},\\
v_{\theta}=v_x \cos{\theta}+v_y \sin{\theta}.
\end{array}
\end{equation}
To lowest order in $q$, the gap function can be expanded as~:
\begin{equation}
\Delta({\bf q}) \simeq
\Delta +(w^{\theta}_3 +\frac{v_{\theta} ^2}{2\Delta}) q^2.
\label{stability}
\end{equation}

\begin{figure}
\includegraphics[height=50mm]{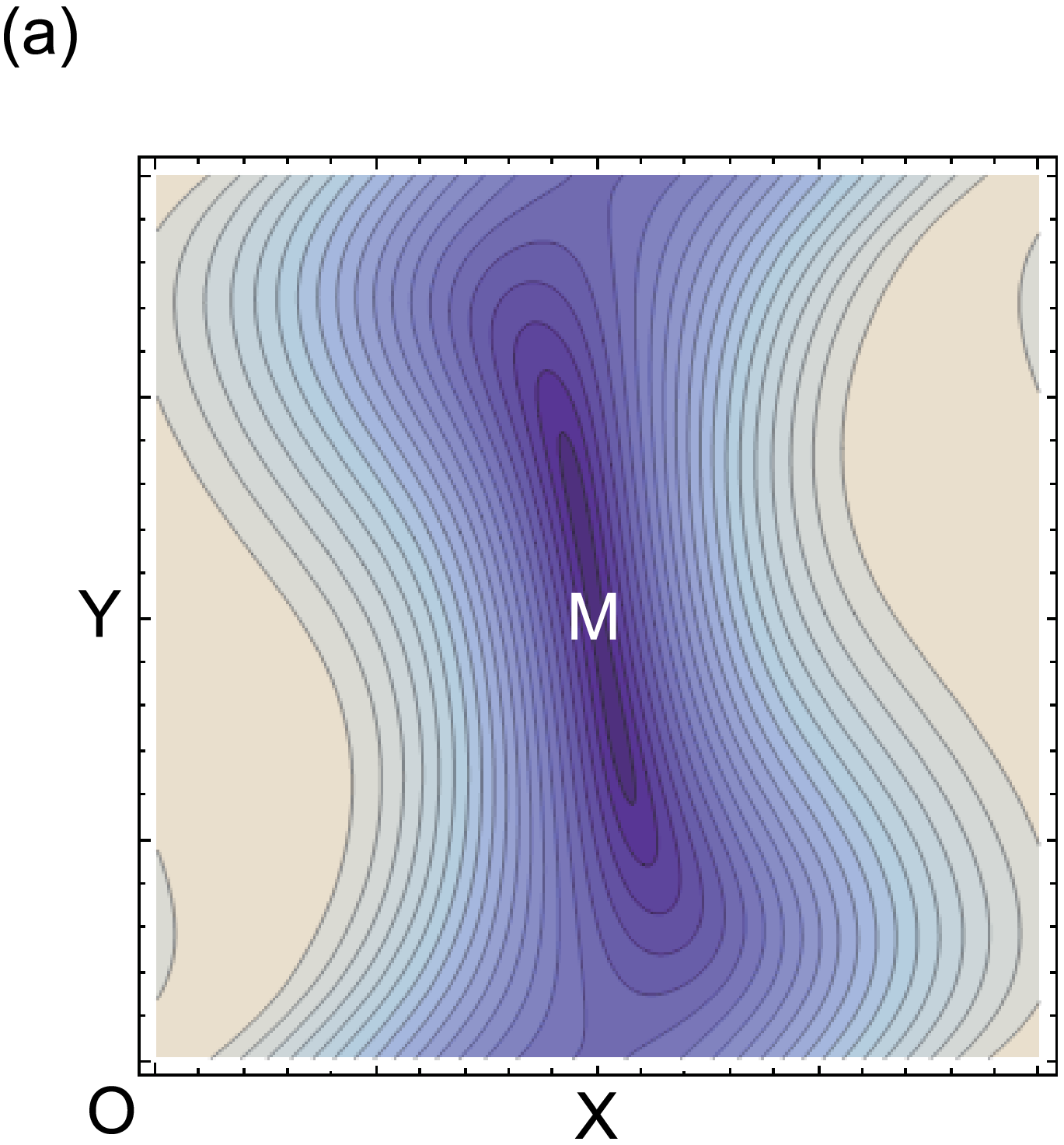}
\includegraphics[height=60mm]{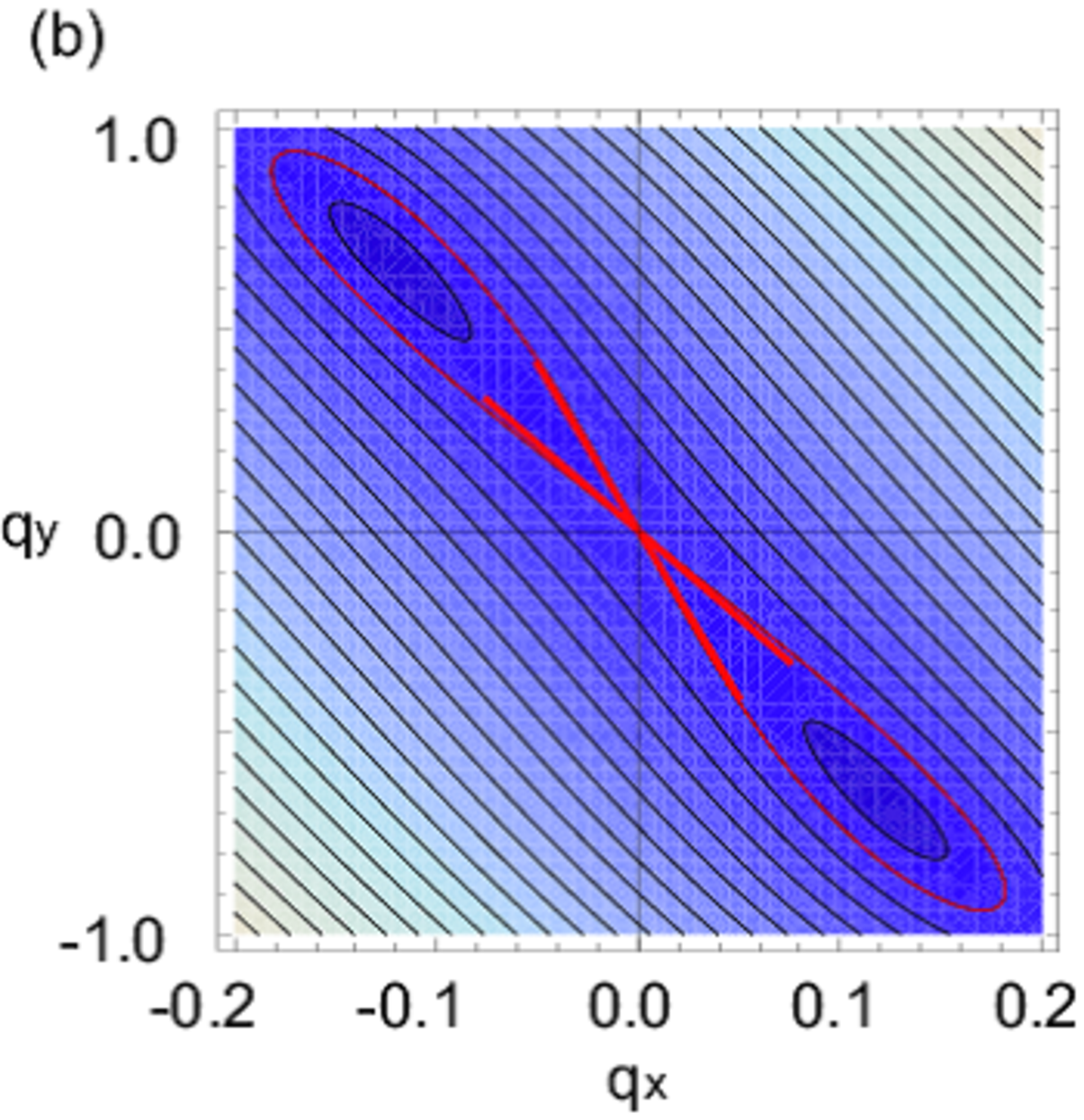}
\caption{\label{fig:merging-bedt}
(a) Contours of the gap function $\Delta_{\bf k}=[\xi_{+}({\bf k}) -\xi_{-}({\bf k})]/2$
 for $P_a = 5.4$ kbar in CO (II) phase
 where  $0 < k_y < 2 \pi$ and
 $0< k_x < 2\pi$.
 The two Dirac points emerge from the M point
  ${\bf k}_M =(\pi,\pm \pi)$.
Points O, X and Y denote (0,0), ($\pi$,0) and (0,$\pi$), respectively.
(b) The same contour $\Delta_{\bf q}$ in the vicinity of the M point, which is taken as the origin (${\bf q} = {\bf k} - {\bf k}_M$). The red curve represents the energy contour $\Delta_{\bf q}=\Delta$. The two red lines denote the angles $\theta_{min}$ and $\theta_{max}$ defined in the text (\ref{thetas}).  Note that on Fig. (b), the scales are different along the two axes.}
\end{figure}

In the CO(II) phase, the function $\Delta({\bf q})$ has a saddle point.
This corresponds to the case where $\textrm{det} S_M  \le 0$ or equivalently
$\overline{w}^{xx} _3\overline{w}^{yy} _3-(\overline{w}^{xy}_3)^2 \le 0$. There is an angular region where $\Delta({\bf q}) < \Delta$.
This happens in an interval $\theta_{min}\le \theta \le \theta_{max}$ such that $(w^{\theta}_3 +\frac{v_{\theta} ^2}{2\Delta})<0$.
  The two angles $\theta_{min},\theta_{max}$,  chosen in the range $[0, \pi]$, are determined by the condition $(w^{\theta}_3 +\frac{v_{\theta} ^2}{2\Delta})=0$, which gives
\begin{equation}
 \tan \theta_{max,min} =
 \frac{-\overline{w}^{xy}_3 \pm
\sqrt{-\textrm{det} S_M} }{\overline{w}^{yy} _3}.
\label{thetas}
\end{equation}
The gap function can be rewritten in terms of these angles as
\begin{eqnarray} \Delta({\bf q}) &=& \Delta + w \sin(\theta - \theta_{max}) \sin(\theta - \theta_{min}) q^2 \; , \nonumber
\\
&=& \Delta +{w \over 2} \left[ \cos\theta_- - \cos 2 (\theta - \theta_+) \right] q^2 \end{eqnarray}
 where
$w$ can be rewritten conveniently as
 \be w= \left[w_M^2 - 4 \textrm{det} S_M\right]^{1/2} \quad , \quad w_M = |\overline{w}_3^{xx} + \overline{w}_3^{yy}| \; . \ee
The angles $\theta_+$ and $\theta_-$
are defined as
\begin{equation} \theta_+={1 \over 2} (\theta_{max} + \theta_{min})  \quad , \quad \theta_-= \theta_{max} - \theta_{min} \; . \end{equation}
They represents respectively the most unstable direction and the angular aperture of the region $\Delta({\bf q}) < \Delta$.
Fig. \ref{fig:merging-bedt}(a) shows
  the gap function  in the CO(II) phase at $P_a = 5.4$ kbar.
Their expression is
\begin{eqnarray}
\theta_+&=&{\pi \over 2}+ {1 \over 2} \arctan \left({ 2 \overline{w}^{xy}_3 \over \overline{w}^{xx}_3 - \overline{w}^{yy}_3 }\right) \; ,
\nonumber \\
\theta_-&=& \arctan {2 \sqrt{-\textrm{det} S_M } \over w_M}  \;, \end{eqnarray}

 Near the merging transition, when $P_a \gtrsim P_M$,   $\theta_+ \rightarrow \theta_M$ where the angle $\theta_M$  given by
\begin{equation}
 \theta_{M}=\pi - \arctan \frac{\overline{w}^{xy}_3}{\overline{w}_3^{yy}}
 \; ,
\label{eq:theta_M}
 \end{equation}
  defines the direction of emergence of the two Dirac points. Moreover $ w \rightarrow w_M = |\overline{w}_3^{xx} + \overline{w}_3^{yy}|$   and  the angular aperture $\theta_-$ vanishes as $ \theta_- \simeq {2 \sqrt{-  \textrm{det} S_M } / w_M }$.

Near the transition, we define a new set of coordinates along the direction of emergence and the perpendicular direction. Writing
 $q'_x = q \cos (\theta -\theta_M)$ and
  $q'_y =  q \sin (\theta -\theta_M)$,
 the gap function can be expanded as

\begin{eqnarray}
\Delta({\bf q})&\simeq & \Delta +  w_M \left( q_y'^2 - {\theta_-^2 \over 4} q_x'^2 \right) \nonumber \\
&\simeq& \Delta  + {\textrm{det} S_M \over  w_M
} q_x'^2 + w_M  q_y'^2   \ .   \end{eqnarray}
This explicitly shows  that $\textrm{det} S_M$ is the single parameter
 for the merging condition.

\begin{figure}
\includegraphics[height=140mm]{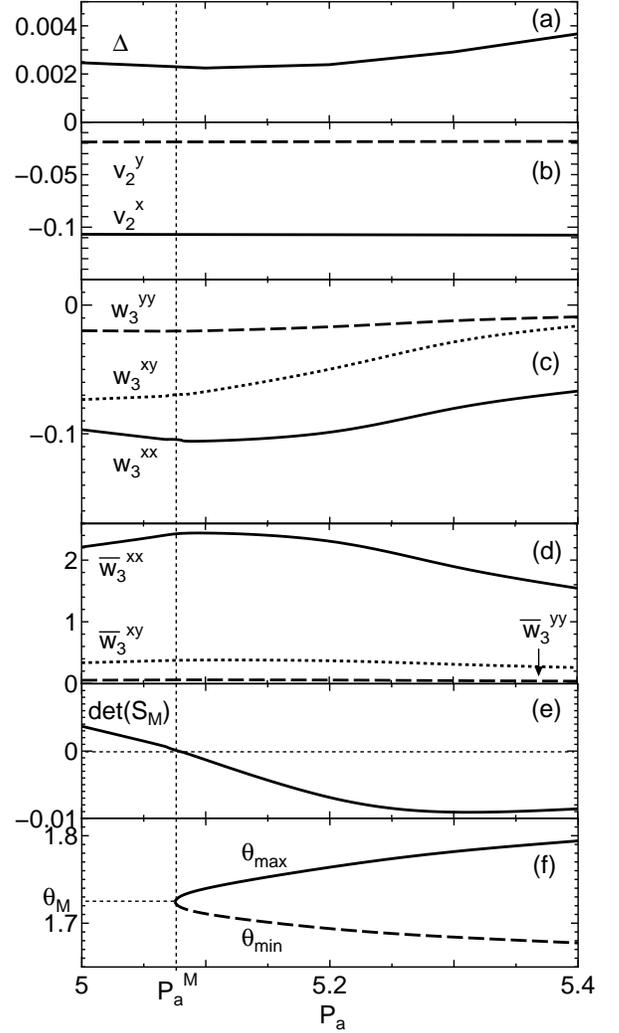}
\caption{\label{EmergingParameter}
Pressure dependence of (a)
$\Delta$, (b) $v_j$, (c) $w_3^{i j}$, (d) $\overline{w}_3^{i j}$, (e)  det$S_M$
 and (f) $\theta_{max}$, $\theta_{min}$
 for $V_a =0.18$ eV  and $V_b =0.05$ eV. The pressure $P_a$ is in kbar.
}
\end{figure}

The $P_a$-dependence of
$\Delta$, $v_j$, $w_3^{i j}$ and  det$S_M$
 for $V_a =0.18$ eV  and $V_b =0.05$ eV are
 shown in Fig.~\ref{EmergingParameter}.
The merging  pressure,  calculated from  $\textrm{det} S_M = 0$, is found to be
$P_a^{\rm M} \simeq$  5.07 kbar, which  almost coincides with that obtained from the direct calculation of energy bands. The merging axis
 $q_x^\prime$-axis is very close to the
$q_y$ axis,
 $\theta_M = 1.73$
as shown in Fig.~\ref{fig:merging-bedt}(a).

%

\subsection{Relation between effective Hamiltonians for $\alpha$-(BEDT-TTF)$_2$I$_3$ and two-component systems}

In this section, we compare the physics of the emergence of Dirac points
in $\alpha$-(BEDT-TTF)$_2$I$_3$, with a possible similar transition in  two-component systems such as graphene or boron nitride (BN) under stress. Near the M-point, the electronic spectrum of these two systems can
 be described by the following Hamiltonian  \cite{Montambaux2009EPJ,Montambaux2009PRB,Fuchs2010}:
\begin{eqnarray}
h_{BN}(\bf q)&=& \left(
         \begin{array}{cc}
           M &  \Delta_{*} + {q_y^2 \over 2 m} - i v_x q_x \\
          \Delta_{*} + {q_y^2 \over 2 m} - i v_x  q_x
           & -M \\
         \end{array}
       \right)  \nonumber \\
       &=&  M \sigma_3 + (\Delta_{*} + {q_y^2 \over 2 m})\sigma_1+ v q_x \sigma_2 \; ,
\label{eq:HBN} \end{eqnarray}
 In graphene $M=0$,  and in boron nitride $M \neq 0$ is proportional to the  energy difference between sites B and N.
  The parameter $\Delta_{*}$ drives the transition from a phase with two Dirac points ($\Delta_{*} <0$) separated from
   $\delta q= 2 \sqrt{-2 m \Delta_{*}}$ to an insulating phase with a gap $\Delta_{*} >0$. This transition has been studied in refs.   \cite{Montambaux2009EPJ,Montambaux2009PRB,Fuchs2010}.

  Here we show that this transition  has the same structure as the transition described above in $\alpha$-(BEDT-TTF)$_2$I$_3$.
To  do this, we have to compare the Luttinger-Kohn Hamiltonian  (\ref{eq:hqtheta}) and the "boron-nitride" Hamiltonian (\ref{eq:HBN}).
This is possible after a rotation of angle $-\varphi$ where
\begin{equation} \tan{\varphi} = \frac{\Delta_{*}}{M} , \end{equation}
from which we obtain a new Hamiltonian:
\begin{eqnarray}
h'_{BN}({\bf q}) =\exp^{{\rm i} \frac{\varphi}{2} \sigma_2} h_{BN}({\bf q})  \exp^{-{\rm i} \frac{\varphi}{2} \sigma_2} \ .
\end{eqnarray}
This Hamiltonian has the form
\begin{eqnarray}
h'_{BN}(\bf q)&=& \left(
         \begin{array}{cc}
         \Delta + w_3 q_y^2 &  w_1 q_y^2 - i v_x q_x \\
        w_1 q_y^2 - i v_x  q_x
           &  \Delta  + w_3 q_y^2 \\
         \end{array}
       \right)  \nonumber \\
       &=&   (\Delta + w_3 q_y^2 ) \sigma_3 + w_1 q_y^2 \sigma_1+ v q_x \sigma_2 \; ,
 \nonumber \\
\label{eq:HBN2} \end{eqnarray}
with the parameters $w_1$, $w_3$ and $\Delta$ given by
\begin{eqnarray}
&& w_1 = {1 \over 2 m }  \cos{\varphi} = {1 \over 2 m } \frac{M }{\sqrt{M^2 + \Delta_{*}^2}} \; , \\
&& w_3 = {1 \over 2 m }   \sin{\varphi} ={1 \over 2 m } \frac{\Delta_{*}}{\sqrt{M^2 + \Delta_{*}^2}},  \\
&& \Delta = \sqrt{\Delta_{*}^2 + M^2}.
\end{eqnarray}

This  Hamiltonian therefore appears as a peculiar case of the general Hamiltonian (\ref{eq:hqtheta}), with the following parameters
$\Delta$, $v_x$, $w_1^{yy}=w_1$, $w_3^{yy}=w_3$ and all other parameters being zero ($w_1^\theta=w_1 \sin^2 \theta$, $w_3^\theta=w_3 \sin^2 \theta$, $v_\theta= v_x \cos \theta$).
In this case, the determinant of the stability matrix is simply  $\textrm{det} S_M= {v_x^2 \over 2 \Delta} w_3$.
The merging transition has the same structure as the one described above.
Here we have simply
\be   \theta_{min}=\arctan  \sqrt{{ v_x^2 \over - 2 \Delta w_3}}= \arctan  \sqrt{{m v_x^2 \over - \Delta_{*}}}  \; ,  \ee
and $\theta_{max}=\pi - \theta_{min}$, so that $\theta_+= {\pi \over 2}$ and
\be    \theta_-= \arctan {2 \sqrt{{- 2 \Delta w_3 \over  v_x^2}} \over \left| 1 +  {2 \Delta  w_3  \over  v_x^2}\right|}=\arctan {2  \sqrt{{- \Delta_{*} \over  m v_x^2}} \over \left|1 +  {  \Delta_{*} \over   m v_x^2}\right|}. \ee

\begin{figure}
\includegraphics[height=60mm]{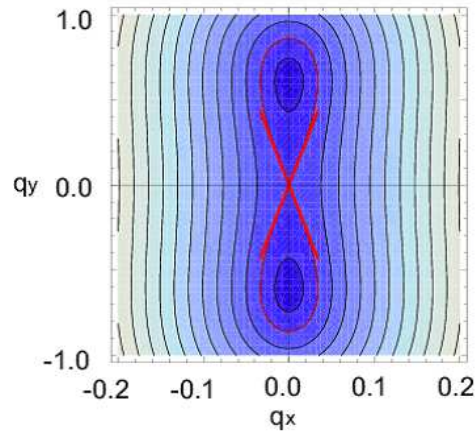}
\caption{\label{fig:merging-nb}
 Contours of $\Delta({\bf q})$ for the BN model. Here we have taken the same parameters as for the CO(I)-CO(II) transition in $\alpha$(BEDT)-TTF)$_2$I$_3$, $\Delta$, $w_1=w_1^{yy}$, $w_3=w_3^{yy}$, $v_x$, other $w_1^{ij}$, $w_3^{ij}$, $v_y$ being zero.   In this simple case, since the parameter $\overline{w}_3^{xy}=0$, the Dirac points stay along the $y$-axis. The red curve corresponds to $\Delta({\bf q})=\Delta$. The two lines indicate the directions $\theta_{max}$ and $\theta_{min}$.}
\end{figure}

In this simple case, the direction of the merging stays constant and perpendicular to the velocity
(see Fig. \ref{fig:merging-nb}). The BN Hamiltonian is interesting since it explicitly reveals the role of the different parameters.    Whereas $w_3$ drives the topological transition, the parameter $w_1$ controls the opening of a gap in both phases. The situation in $\alpha$-(BEDT-TTF)$_2$I$_3$ is more involved since there are 9 parameters but   the main features of the topological transition can be understood within a simplified Hamiltonian with only 4 parameters.

\section{Berry curvature  associated with the topological transition}

In this section we show that the emergence of the Dirac  pair is well
characterized by the appearance of a non trivial Berry  curvature ${B}_{1 \uparrow} ({\bf k})$ for the conduction band.
This Berry curvature, which plays the role of a ${\bf k}$ dependent effective magnetic field
in the Brillouin zone, appears to be sharply peaked around the two massive Dirac points (see Fig.~\ref{Berry}(b)).
We then show how to associate a Berry phase to each Dirac point.

\subsection{Full Brillouin zone computation of the Berry curvature within the four-band model}

 For a multiband system, the Berry curvature ${B}_n ({\bf k})$ of the $n^{\textrm{th}}$ band is
given by \cite{Berry}:
\begin{equation}
{\bf B}_n ({\bf k})={\rm rot}_{\bf k} {\bf A}_n ({\bf k})\; .
\label{eq:eq24}
\end{equation}
where ${\bf A}_n ({\bf k})$ denotes the so-called Berry connection and is written as
\begin{equation}
{\bf A}_n ({\bf k})=-{\rm i} <n ({\bf k}) \vert \partial_{\bf k}
\vert n ({\bf k})>,
\end{equation}
 and $\vert n ({\bf k})>$ is an eigenvector
 of Eq.~(\ref{eigenvalue}).

For our model Hamiltonian (\ref{eq:MF_Hamiltonian}) which is
diagonal in spin index and with non degenerate bands $\xi_{n \sigma}({\bf k})$
that never cross for any ${\bf k}$, the Berry curvature ${\bf B}_{n \sigma} ({\bf k})\equiv {B}_{n \sigma} ({\bf k}) {\bf u}_z$
can be computed from
\begin{equation}
\label{Bn}
B_{n \sigma} ({\bf k})=-{\rm i} \sum_{m \ne n} \frac{v_{n m \sigma}^x({\bf k})v_{mn \sigma}^y({\bf k})+c.c.}
{(\xi_{ n \sigma} ({\bf k}) -\xi_{m \sigma} ({\bf k}) )^2},
\end{equation}
 where
\begin{equation}
\begin{array}{ll}
v_{n m \sigma}^{x,y}({\bf k})&=
<n \sigma ({\bf k}) \vert \partial_{k_{x,y}} H_{\sigma} ({\bf k})
 \vert m \sigma ({\bf k})>\\
&=\sum_{\alpha \beta} d_{n \alpha \sigma} ({\bf k})^* d_{m \beta \sigma} ({\bf k}) \partial_{k_{x,y}} \epsilon_{\alpha \beta} ({\bf k}) \; ,
\end{array}
\end{equation}
are the $nm$ interband matrix elements of the velocity operator.
 $n,m=1,2,3,4$ are  Bloch bands indices
and $\alpha=A$, $A^\prime$, $B$, $C$ are  sites in $\alpha$-(BEDT-TTF)$_2$I$_3$.
Note that since $v_{n m \sigma}^{x,y}({\bf k})=\left(v_{m n\sigma}^{x,y}({\bf k})\right)^*$, we immediately obtain that
$B_{n \sigma} ({\bf k})=-\sum_{m \ne n} B_{m \sigma} ({\bf k})$. Therefore computing $B_{1 \sigma} ({\bf k})$
fully characterizes our $3/4$ filled system.

\begin{figure}[htbp]
 \begin{minipage}{0.5\hsize}
  \begin{center}\leavevmode
   \hspace*{0cm}\includegraphics[width=7cm]{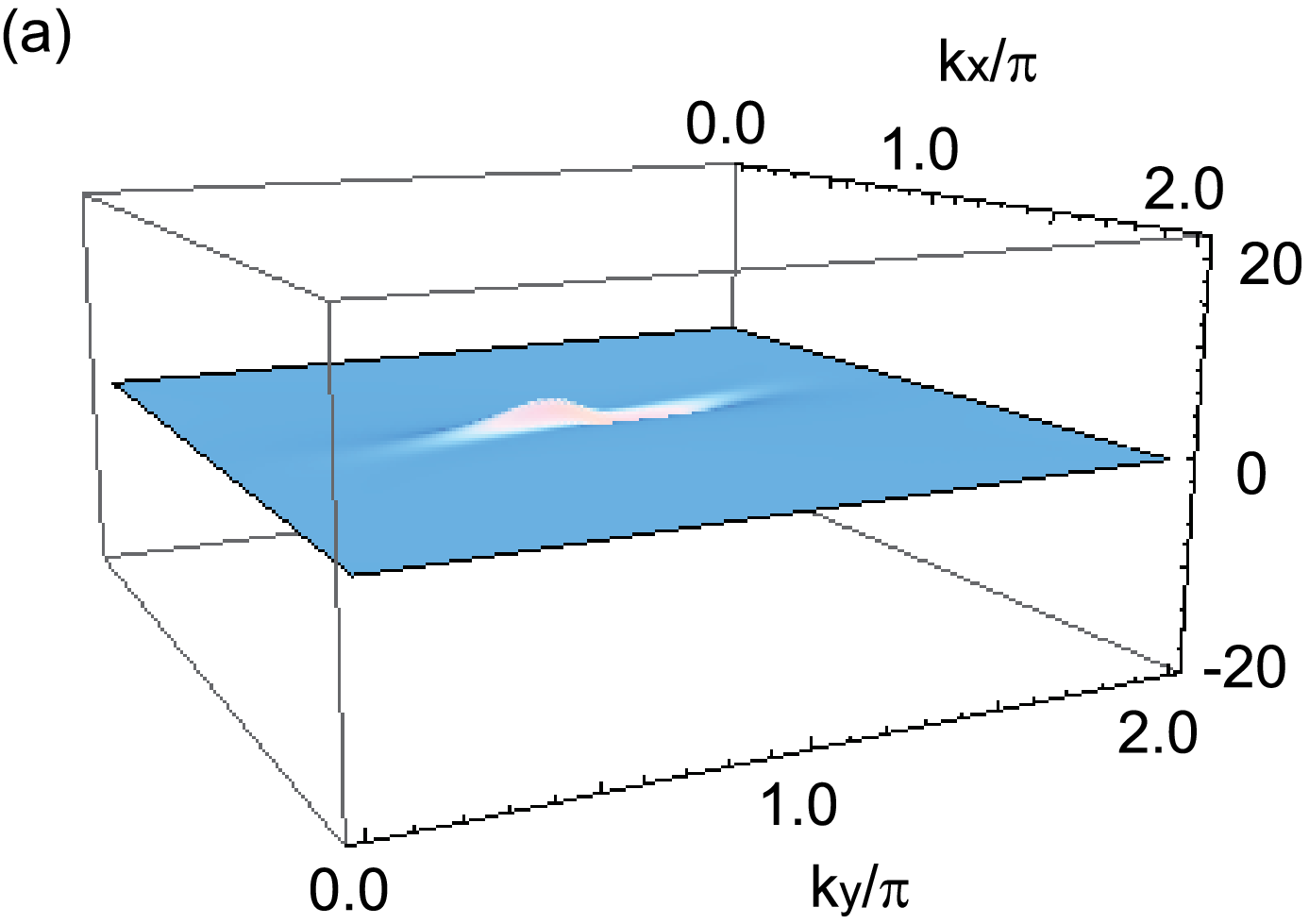}
  \end{center}
 \end{minipage}
 \begin{minipage}{0.5\hsize}
  \begin{center}\leavevmode
   \hspace*{0cm}\includegraphics[width=7cm]{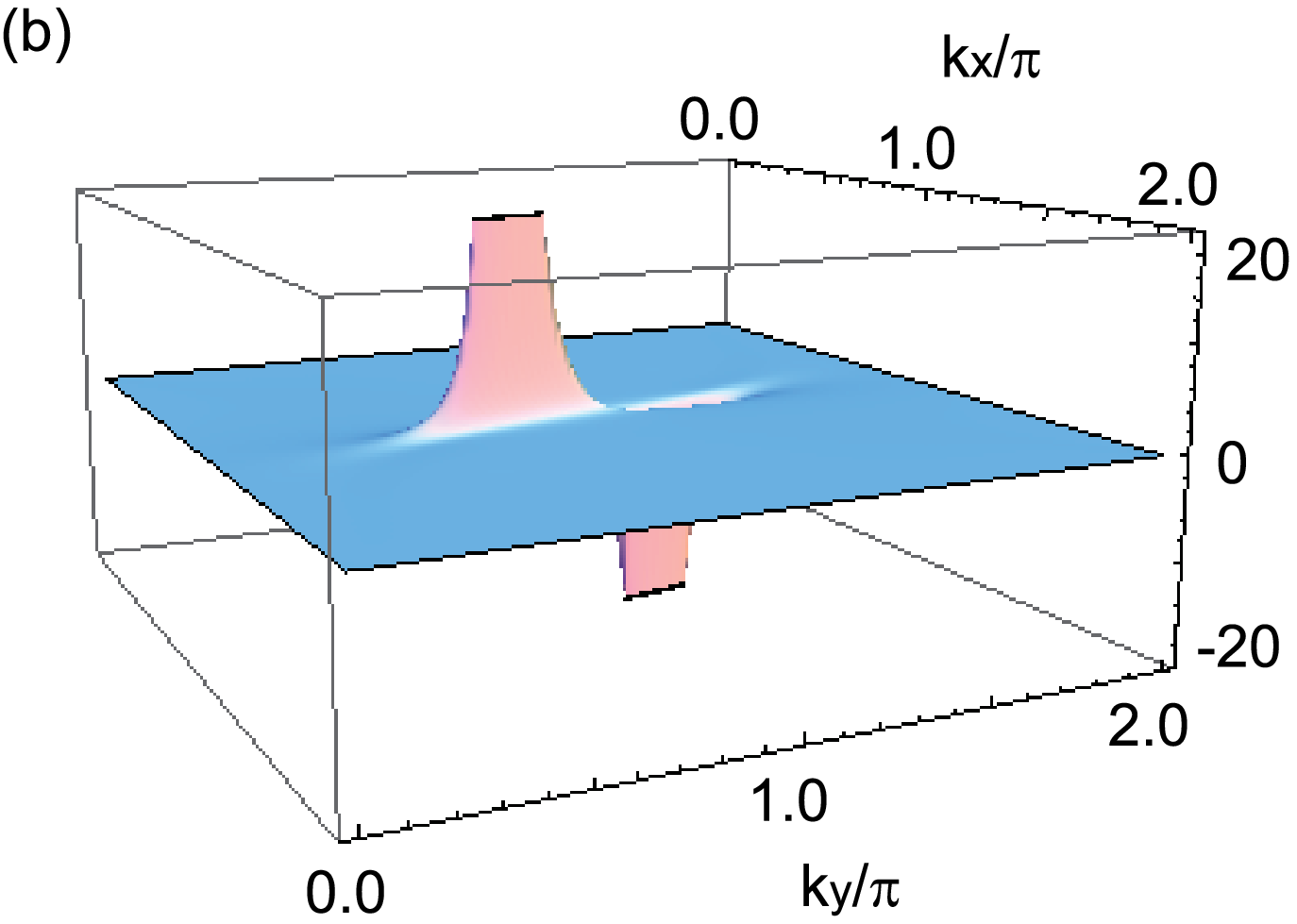}
  \end{center}
 \end{minipage}
 \caption{
(a) Berry curvature $B_{1 \uparrow} ({\bf k})$ in CO(I)
with $P_a =4.4$ kbar,
(b) Berry curvature $B_{1 \uparrow} ({\bf k})$ in CO(II)
with $P_a =5.4$ kbar, where $V_a =0.18$ eV and $V_b =0.05$ eV.
 The center denotes the M point.
}
 \label{Berry}
\end{figure}

Fig. \ref{Berry} shows the typical result for $B_{1 \uparrow} ({\bf k})$ in the insulating
CO(I) ($P_a =4.4 $ kbar) and CO(II) ($P_a=5.4$ kbar) phases.
In the CO(II) phase, there is a pair of massive Dirac particles and consequently the Berry curvature
is sharply peaked around the position of the Dirac points ${\bf k_{\pm}}$.
The peaks at ${\bf k_{\pm}}$ are strongly anisotropic and have the same magnitude but opposite sign, reflecting the so-called opposite
chirality associated to the two Dirac particles.
Conversely, in the CO(I) phase,  the intensity of the Berry curvature
becomes small owing to cancelation of the positive and negative peaks.

\subsection{Berry curvature of the effective two bands Luttinger-Kohn model}

In order to understand in a more quantitative way these numerical results, we show here how the Berry curvature ${B}_{1 \uparrow} ({\bf k})$ can be computed  directly from the low energy effective
$2 \times 2$ Luttinger-Kohn reduced Hamiltonian
$H_{\uparrow}^{KL}({\bf k_M+q})=f_0({\bf q})\sigma_0+ {\bf f}({\bf q} )\cdot {\boldmath \sigma}$. ${\bf f}$ is the vector of components $(f_1, f_2,f_3)$ (Eq. \ref{KLhamiltonian}).
For this $2\times2$ Hamiltonian, the Berry curvature $B_{1\uparrow} ({\bf q}={\bf k-k_M})$
(in cartesian and polar coordinates) reads
\begin{equation}
\begin{array}{ll}
B_{1\uparrow} ({\bf q})&=\dfrac{(\partial_{k_x} {\bf f} \times \partial_{k_y} {\bf f}) \cdot  {\bf f}}{2 {\Delta^3({\bf q})}}\\
&=\dfrac{(\partial_{q} {\bf f} \times \partial_{\theta} {\bf f}) \cdot  {\bf f}}{2 q  {\Delta^3({\bf q})}}.
\end{array}
\end{equation}
From Eq.(\ref{eq:hqtheta}) we obtain:
\begin{equation}
B_{1\uparrow}(q,\theta)=\frac{\Delta (2w_1^{\theta}{v_{\theta}} '-v^{\theta}{w_1 ^{\theta}}')q +v_\theta(w_3^{\theta}{w_1 ^{\theta}}'-w_1^{\theta}{w_3 ^{\theta}}')q^3  }
{2\left[  (\Delta  +w_3^\theta q^2 )^2 +v_\theta^2 q^2 +(w_1^\theta)^2 q^4                                      \right]^{3/2}},
\label{Curvature_analytical}
\end{equation}
where we have defined the derivatives
\begin{equation}
\begin{array}{l}
{w^{\theta} _1} '=(w^{yy} _1-w^{xx}_1) \sin{2 \theta}+ 2 w^{xy} _1 \cos 2{\theta},\\
{w^{\theta} _3}'=(w^{yy} _3-w^{xx}_3) \sin{2 \theta}+2 w^{xy} _3  \cos2{\theta},\\
{v_{\theta}}'=-v_x \sin{\theta}+v_y \cos{\theta}.
\end{array}
\end{equation}
The expression Eq.(\ref{Curvature_analytical}) shows that the $w_{1} ^{\theta}$ F term is essential
to have a non-vanishing Berry curvature.
We note that this expression is quite general. In particular, for the boron nitride Hamiltonian described in previous section, we obtain
\begin{equation}
\label{Curvature_analytical2}
B^{BN} ({\bf q}) =\frac{ -\Delta  w_1 v_x q_y}
{((\Delta  +w_3 q_y^2 )^2 +v_x^2 q_x^2 +w_1^2 q_y^4)^{\frac{3}{2}}}.
\end{equation}
The $q^3$ term in the numerator (that determines the large $q$ tail of the Berry curvature)
vanishes here because $w_{1} ^{\theta}$ and  $w_{3} ^{\theta}$ have identical $\theta$ dependences.

\subsection{Berry phase associated to Dirac points}

The integral of the Berry curvature $B_{1 \uparrow}({\bf k})$  over the full  Brillouin zone (BZ) is a topological quantity called a Chern number \cite{Chern}
\begin{equation}
\label{eq:Chn}
Ch_{1 \uparrow} =\frac{1}{2 \pi} \int_{BZ} {\it d}S B_{1 \uparrow} ({\bf k}).
\end{equation}
In our system, $Ch_{1 \uparrow}=0$ since the curvatures associated to the two Dirac points exactly compensate.
In order to characterize the contribution of each Dirac point, we define a Berry phase which is the integral over
 an appropriate region in ${\bf k}$ space around a
given Dirac point as explained in \cite{Fuchs2010}:
\begin{figure}
\includegraphics[height=70mm]{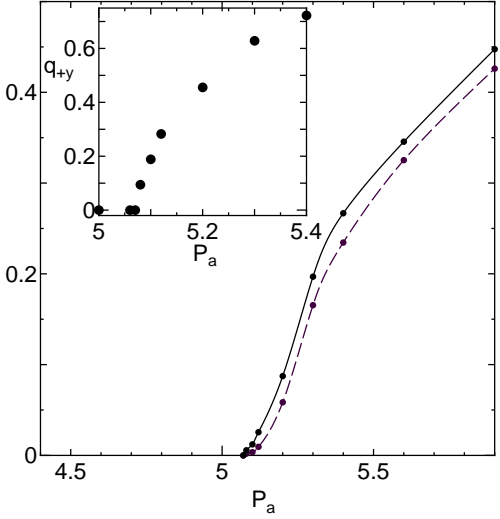}
\caption{
\label{ChernNumber}
 $P_a$-dependence of the Berry phase $|\Gamma({\bf k}_\pm)|$ defined in
 Eq.~(\ref{eq:Ch_analytical})
(solid line)  compared with the  expression $\Gamma_\pm$ given in
 Eq.~(\ref{Chrm})  (dashed line).
The inset denotes ${q_+}_y$,  the y-component of one Dirac point
 ${\bf k}_+$,  where ${\bf q} = {\bf k} - {\bf k}_M$.
}
\end{figure}
\begin{equation}
\label{eq:Ch_analytical}
\Gamma({\bf k}_{\pm})  =\frac{1}{2 \pi} \int_{S({\bf k}_{\pm})} {\it d}S B_{1\uparrow} ({\bf k}) \; .
\end{equation}
The area  $S({\bf k}_{\pm})$  denotes the  region around a given Dirac point ${\bf k}_{\pm}$ where $\Delta({\bf k})$  contours are closed ($\Delta({\bf k}) < \Delta$). This is the region inside the red curve
in Figures \ref{fig:merging-bedt} and \ref{fig:merging-nb}.

Here we compare  $\Gamma({\bf k}_{\pm})$  with the following expression $\Gamma_{\pm}$
\begin{equation}
\Gamma_{\pm}
=\mp \frac{1}{2} \left(1 - \frac{\Delta({\bf k}_{\pm})}{\Delta({\bf k}_{M})} \right) \;.
 \label{Chrm}
\end{equation}
This expression was obtained in Ref.~\onlinecite{Fuchs2010} for a general class of $2 \times 2$ Hamiltonians that include the BN model for which we find
\begin{equation}
\Gamma_{\pm}
= \mp \frac{1}{2} \left(1 - \frac{w_1}{\sqrt{w_1^2+w_3^2}} \right)  = \mp \frac{1}{2} \left(1 - \frac{M}{\sqrt{M^2+\Delta_*^2}} \right) \;.
 \label{Chrm2}
\end{equation}
Therefore in the vicinity of the transition ($w_3 \rightarrow 0_-$), $\Gamma_\pm$ varies as
\begin{equation} |\Gamma_\pm| \simeq {1 \over 4}  {w_3^2 \over w_1^2} \ . \label{Gammaw3} \end{equation}
At the same time, the separation $\delta q$ between Dirac points
varies as
 \begin{equation} \delta q \simeq   {(- \Delta w_3)^{1/2} \over w_1}  \ . \label{deltaqw3}  \end{equation}
\noindent
 The $P_a$-dependence of $\Gamma({\bf k}_{\pm})$ is compared with
that of  $\Gamma_{\pm}$  in
  Fig.~\ref{ChernNumber}.
Although the behavior of  $\Gamma({\bf k}_{\pm})$ is  qualitatively similar to
 $\Gamma_{\pm}$,
 there is
 a slight difference between these two quantities, which may
come from the effect of the other bands    in Eq.~(\ref{Bn}).
This multi-band effect
  is discussed in  a separate paper \cite{Suzumura_Berry}.
 Both quantities $\Gamma({\bf k}_{\pm})$ and $\Gamma_{\pm}$ vanish below a critical pressure which coincides with the merging pressure   $P_a^M$ found in  Fig~\ref{EmergingParameter},
 within numerical accuracy.
This is consistent with  the effective Hamiltonian which leads to   $|\Gamma({\bf k}_{\pm})| \propto (P_a - P_a^{\rm M})^2$ and
$q_{+y} \propto (P_a - P_a^{\rm M})^{1/2}$. This is also seen quite simply in the BN model for which we obtain, from Eqs.~(\ref{Gammaw3}) and (\ref{deltaqw3}), $|\Gamma_\pm| \propto w_3^2 \propto (\textrm{det}S_M)^2$ and $q_{+y} \propto w_3^{1/2} \propto (\textrm{det}S_M)^{1/2}$.

\section{Summary and discussion}

In this paper,
  we re-examined the electronic properties
  in the stripe charge ordered states in the organic conductor $\alpha$-(BEDT-TTF)$_2$I$_3$.
With increasing pressure,  a transition occurs from
 an insulating charge ordered (CO) state to a metallic charge ordered (COM) state.
 In this work, we have found a new topological transition which further splits each of the CO and COM phases into two phases, respectively CO(I)-CO(II) and COM(I)-COM(II). We specifically consider the CO(I)-CO(II) transition which separates a phase (CO(I)) with a usual gap at the M-point of the Brillouin zone from a phase (CO(II)) where the gap exhibits a double minimum with a local maximum at the M-point. This transition corresponds to the emergence of a  pair of Dirac points.

This modification of the band structure is described
      by an effective $2 \times 2$ low energy Luttinger-Kohn Hamiltonian with 9 parameters that can be extracted from a numerical Hartree calculation. From a detailed study of this Hamiltonian and the corresponding energy spectrum, we show that the transition is driven by  a single quantity
  $\textrm{det} S_M$ (Eq.~(\ref{eq:singleparameter})) which is an appropriate combination of these parameters and whose sign changes at the transition.
A similar scenario occurs for the COM(I)-COM(II) transition. \GMFP{ The Hartree contribution, which induces on-site potentials, is crucial to break inversion and time-reversal symmetries, leading to the CO state. The exchange (Fock) term modifies the hopping energies but does not break the symmetry, so that it is not responsible for the CO state. It may only modify the relative stability of the different phases. Moreover, in the ZGS  phase, it does not disturb the Dirac spectrum as shown by first principle calculations$^{13,14}$. Similarly we believe that going beyond the mean field approximation does not change qualitatively the topological transition around the M point  but may only modify the parameters of the Luttinger-Kohn Hamiltonian.}

We compare the structure of this transition with a similar situation which occurs in a simple model for boron-nitride \cite{Montambaux2009EPJ,Montambaux2009PRB,Fuchs2010}, where the emergence of a pair of Dirac points is quite comparable,
although the driving forces are different.

The existence of a pair of Dirac points is characterized by a special structure of the
 the Berry curvature inside the Brillouin zone.
In  the CO(II) and COM(II) states,
 the latter    shows two sharp peaks  with opposite signs.
On the other hand, in CO(I) and COM(I) phases,  the Berry curvature
  becomes very small owing to cancelation of the positive
   and negative contributions.
 The existence of the Dirac point is also
 verified by integrating the  Berry curvature
  over a region limited by  a closed energy contour  around a single point.

It would be interesting to observe directly the modification of the energy spectrum associated to the topological transition by  angle-resolved photoelectron spectroscopy experiments. Moreover  this topological transition could be probed in a magnetic field by the modification of the Landau level structure \cite{Montambaux2009EPJ}, therefore by {\it e.g.}  magnetoresistance \GMFP{and Hall} experiments, \GMFP{in a range of pressure which is accessible experimentally}.
Finally we mention the existence of  another conductor,  $\alpha$-(BETS)$_2$I$_3$ which is  also a good  candidate for a Dirac-like electron spectrum
\cite{TakahashiBETS}. Moreover, since the BETS molecule contains relatively heavy atoms,
the spin-orbit interaction may be effective leading to non-zero Chern number as shown in the Weyl Hamiltonian with
the spin-orbit interaction \cite{Kane-Mele2005}.

\acknowledgements
The authors are thankful to J.-N. Fuchs and M. O. Goerbig for fruitful discussions.
Y.S. is indebted to the Daiko foundation for the financial aid
 in the present work.
This work was financially supported in part
 by Grant-in-Aid for Special Coordination Funds for Promoting
Science and Technology (SCF), Scientific Research on Innovative
Areas 20110002, and was also financially supported by a Grant-in-Aid for Special Coordination Funds
for Promoting Science and Technology (SCF) from the Ministry of Education, Culture, Sports, Science
and Technology in Japan,
and Scientific Research Grant No. 19740205, Grant No. 22540366, and Grant No. 23540403 from the Ministry of
Education, Culture, Sports, Science and Technology in Japan. This work was partially supported by the NANOSIM-GRAPHENE project (ANR-09-NANO-016-01) of ANR/P3N2009.

\appendix
\section{Band Structure in the stripe Charge Ordered state}

\begin{figure}
\includegraphics[height=70mm]{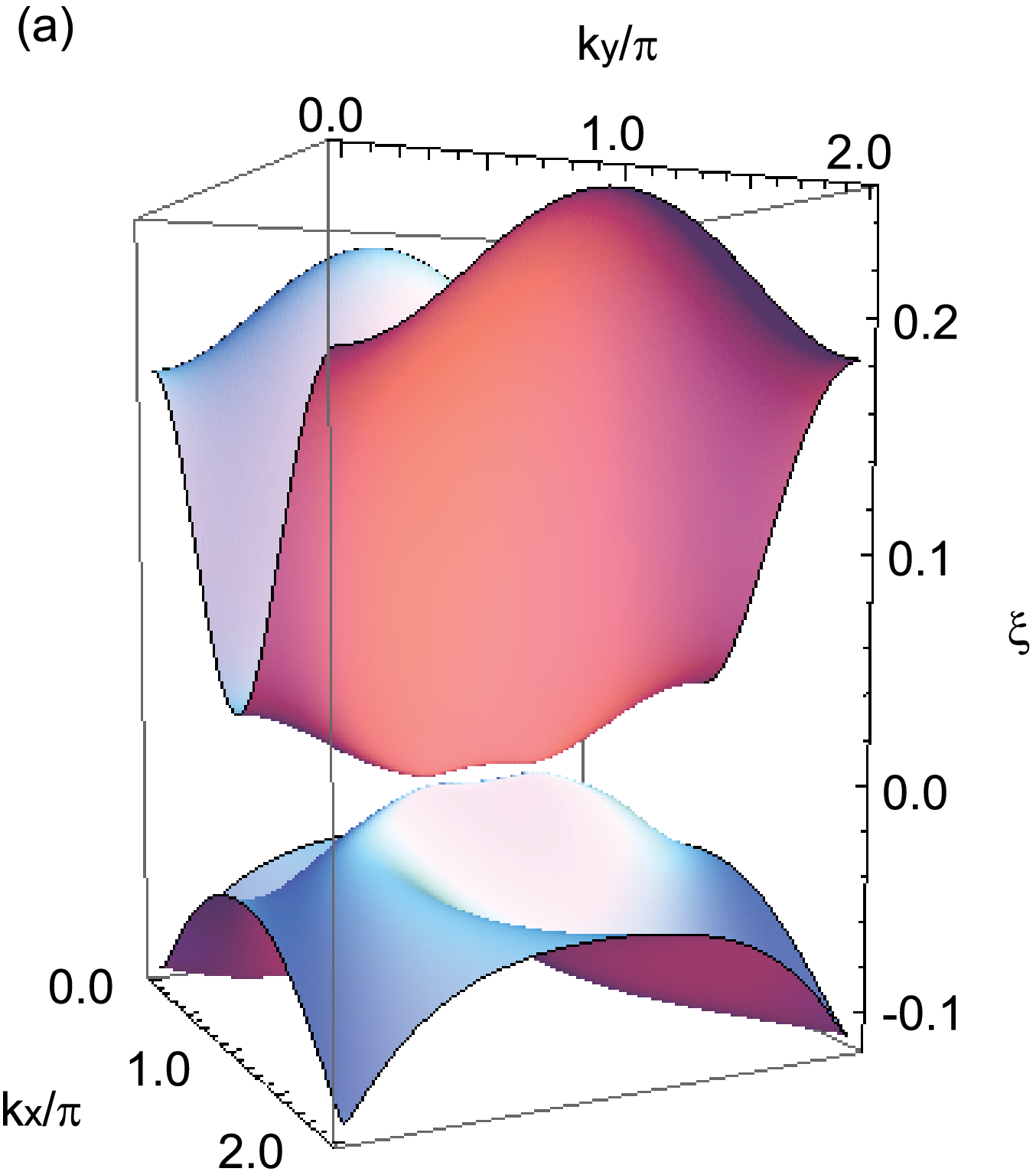}
\includegraphics[height=70mm]{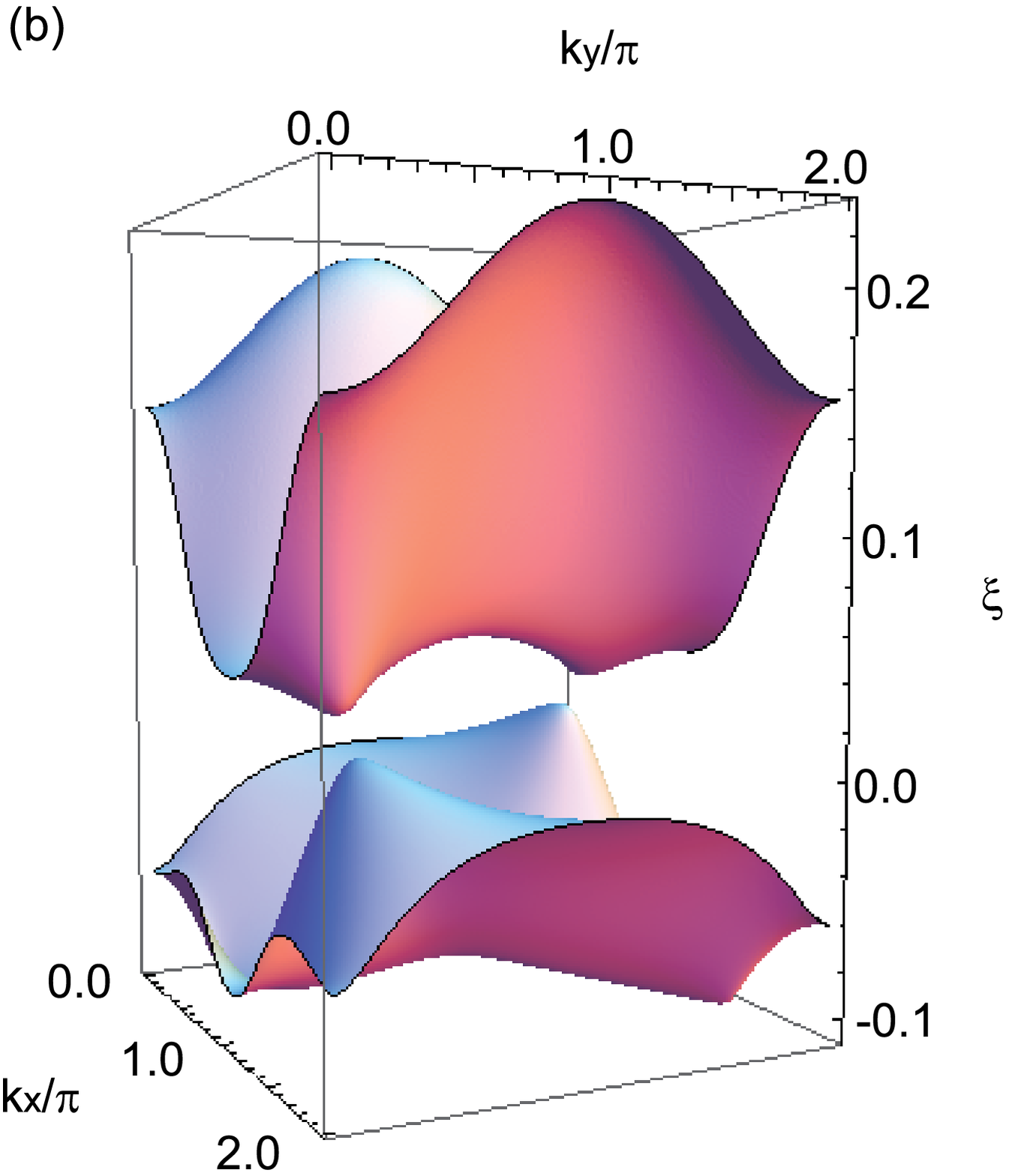}
 \caption{The energy bands  of
   $\xi_{1 \sigma}$ and $\xi_{2 \sigma}$ for
  $\uparrow$   spin band (a) and the   $\downarrow$ spin band (b)
  in the CO(II) for $P_a =5.4$kbar, $V_a =0.18$ eV and $V_b =0.05$ eV.
 The center denotes the M point.
}
 \label{CO(II)Bands}
\end{figure}

In the stripe CO state,
 the energy band with  $\uparrow$ spin is different from
 that with  $\downarrow$ spin  owing to inversion symmetry breaking and time reversal symmetry breaking.
For the CO(I)  phase
 ( $V_a$ = 0.17 eV, and  $P_a$ = 3 kbar in Fig.~\ref{PhaseDiagram})
\cite{Kobayashi2008SCMC},
 the minimum point of the energy gap is located at
${\bf k}0_{\uparrow} = {\bf k}_M =
(\pi, \pm \pi )$ (the M-point )  for $\xi_{1 \uparrow}({\bf k})$,
   and at ${\bf k}0_{\downarrow} =\pm (0.36 \pi , -0.62 \pi)$ for $\xi_{1 \downarrow}({\bf k})$.
 The $\uparrow$ spin  band
   has a small gap $\Delta_{\uparrow} =0.0033$ eV at the M-point and  is very anisotropic in its vicinity.
The  $\downarrow$ spin band has a large gap $\Delta_{\downarrow} =0.0171$ eV
 and is more isotropic.
In the CO(II) phase (Fig.~\ref{PhaseDiagram}),
on the other hand,
there is a pair of minima of the gap in the $\uparrow$ spin band
as shown in Fig.~\ref{CO(II)Bands}.
The gap-minima  of the  $\uparrow$ spin band are
  located at ${\bf k}0_{\uparrow} =\pm (0.95 \pi , -0.71 \pi )$ for $P_a =5.4$kbar, $V_a =0.18$ eV and $V_b =0.05$ eV.

 The energy dispersion in COM(II) is similar to that  of CO(II).
However, for COM(II),
the Fermi level is  located in the conduction band for the  $\uparrow$ spin band,
  and  in the valence band for the $\downarrow$ spin band
(Fig \ref{DiagramBand} d).
When the pressure is decreased, those two minimum points of the  $\uparrow$ spin bands merge at the M-point
 while the quantitiy $\xi_{1 \uparrow} - \xi_{2 \uparrow}$
 ( $\xi_{1 \downarrow} - \xi_{2 \downarrow}$ )
  remains finite on the boundary line between
    COM(I) and COM(II).

\begin{thebibliography}{}
%
\bibitem{TajimaRev2009}
 N.\ Tajima and  K.\ Kajita K,
   Sci.\ Tech.\ Adv.\ Mater. \textbf{10}, 024308 (2009).
\bibitem{SeoRev2004}
H.\ Seo,  C.\ Hotta, and  H. Fukuyama H,
   Chem.\ Rev.\ \textbf{104}, 5005 (2004).
\bibitem{KobayashiRev2009}
A.\ Kobayashi S.\ Katayama, and Y.\ Suzumura,
   Sci.\ Tech.\ Adv.\ Mater. \textbf{10}, 024309 (2009).
\bibitem{Kino-Fukuyama}
H.\ Kino and H.\ Fukuyama,
  J.\ Phys.\ Soc.\ Jpn.\ \textbf{64}, 4523 (1995).
\bibitem{Seo}
H.\ Seo,
   J.\ Phys.\ Soc.\ Jpn.\ \textbf{69}, 805 (2000).
\bibitem{Hotta}
C.\ Hotta,
  J.\ Phys.\ Soc.\ Jpn.\ \textbf{72}, 840 (2003).
\bibitem{TakahashiStripe}
T. Takahashi,
 Synth.\ Met.\ \textbf{133-134}, 26 (2003).
\bibitem{Tajima-Ebina2002}
N.\ Tajima, A.\ Ebina-Tajima, M.\ Tamura,
 Y.\ Nishio, K.\ Kajita,
  J.\ Phys.\ Soc.\ Jpn.\ \textbf{71}, 1832 (2002).
\bibitem{Kobayashi2005}
A.\ Kobayashi, S.\ Katayama, and Y.\ Suzumura,
 J.\ Phys.\ Soc.\ Jpn.\ \textbf{74}, 2897 (2005).
\bibitem{KajitaFirst}
K.\ Kajita, T.\ Ojiro, H.\ Fujii, N.\ Nishio,
 H.\ Kobayashi, A.\ Kobayashi, R.\ Kato,
 J.\ Phys.\ Soc.\ Jpn.\ \textbf{61}, 23 (1992).
\bibitem{Katayama2006ZGS}
S.\ Katayama, A.\ Kobayashi, and Y.\ Suzumura,
 J.\ Phys.\ Soc.\ Jpn.\ \textbf{75}, 054705 (2006).
\bibitem{Kondo2005}
R.\ Kondo S.\ Kagoshima, and J.\ Harada,
 Rev.\ Sci.\ Instrum.\  \textbf{76}, 093902 (2005).

\bibitem{Kino}
H. Kino and  T. Miyazaki,
 J.\ Phys.\ Soc.\ Jpn.\ \textbf{75}, 034704 (2006).
\bibitem{Ishibashi}
S. Ishibashi, T. Tamura, M. Kohyama, and K. Terakura,
 J.\ Phys.\ Soc.\ Jpn.\ \textbf{75}, 015005 (2006).
\bibitem{Kobayashi2007}
A. Kobayashi, S. Katayama, Y. Suzumura, and H. Fukuyama,
 J.\ Phys.\ Soc.\ Jpn.\ \textbf{76}, 034711 (2007).
\bibitem{Montambaux2008TiltedWeyl}
M.O. Goerbig, J.-N. Fuchs,  G. Montambaux, and  F. Pi$\acute{{\rm e}}$chon,
  Phys.\ Rev.\ B \textbf{78 }, 045415 (2008).
\bibitem{Kobayashi2008}
A. Kobayashi, Y. Suzumura, and H. Fukuyama,
 J.\ Phys.\ Soc.\ Jpn.\ \textbf{77}, 064718 (2008).
\bibitem{Tajima2009}
N. Tajima,  S. Sugawara,  R. Kato,  Y. Nishio, and  K. Kajita,
 Phys.\ Rev.\ Lett.\  \textbf{102}, 176403 (2009).
\bibitem{Morinari2009}
K. Morinari,  T. Himura, and  T. Tohyama,
 J.\ Phys.\ Soc.\ Jpn.\ \textbf{78},  023704 (2009).
\bibitem{Dietl}
P. Dietl, F. Pi$\acute{{\rm e}}$chon, and Montambaux G
 Phys.\ Rev.\ Lett.\  \textbf{100}, 236405 (2008).
\bibitem{Montambaux2009EPJ}
G. Montambaux,  F. Pi$\acute{{\rm e}}$chon,  J.-N. Fuchs, and   M.O. Goerbig,
 Eur.\ Phys.\ J.\ B \textbf{72}, 509 (2009).
\bibitem{Montambaux2009PRB}
G. Montambaux, F. Pi$\acute{{\rm e}}$chon,  J.-N. Fuchs, and
M.O. Goerbig,
  Phys.\ Rev.\ B \textbf{80}, 153412 (2009).
\bibitem{Berry}
M.\ V.\ Berry,
 Proc.\ Roy.\ Soc.\ London\ A\
\textbf{392},   45 (1984).
\bibitem{Fuchs2010}
 J.-N. Fuchs,   F. Pi$\acute{{\rm e}}$chon,
 M.O. Goerbig, and  G. Montambaux,
 Eur.\ Phys.\ J.\ B \textbf{77}, 351 (2010).
\bibitem{Mori1984}
T. Mori, A. Kobayashi, Y. Sasaki, H. Kobayashi, G. Saito,  and
  H.Inokuchi,
 Chem.\ Lett.\  957 (1984).
\bibitem{Mori1999}
T. Mori, H. Mori, and S. Tanaka,
 Bull.\ Chem.\ Soc.\ Jpn.\  \textbf{72},  179 (1999).
\bibitem{Kakiuchi2007}
T. Kakiuchi, Y. Wakabayashi, H. Sawa, T. Takahashi, and T. Nakamura,
 J.\ Phys.\ Soc.\ Jpn.\ \textbf{76},  113702 (2007).
\bibitem{Takahashi2008CD}
Y. Takano, K. Hiraki, T. Takahashi, and H.M. Yamamoto,
 J.\ Phys.\ Soc.\ Jpn.\ \textbf{79},  104704 (2010).
\bibitem{Kobayashi2004_11}
A. Kobayashi, S. Katayama, K. Noguchi, and Y. Suzumura,
 J.\ Phys.\ Soc.\ Jpn.\ \textbf{73}, 3135 (2004).
\bibitem{Kobayashi2008SCMC}
A. Kobayashi, S. Komaba, S. Katayama, and Y. Suzumura,
 J.\ Phys.,\ Conf.\ Ser.\ \textbf{132},  012002 (2008).
\bibitem{LuttingerKohn}
J.M. Luttinger  and W. Kohn W
  Phys.\ Rev.\ \textbf{97}, 869 (1955).
\bibitem{Katayama_EPJBS}
S. Katayama, A. Kobayashi, and Y. Suzumura,
 Eur.\ Phys.\ J.\ B \textbf{67}, 139 (2009).
\bibitem{remark1} The possibility of a transition towards a local maximum is also possible, but does not correspond to the  physical situation discussed in this paper.
\bibitem{Chern}
 D.J. Thouless,   {\it Topological Quantum Numbers in Nonrelativistic
Physics} (World Scientific, Singapore, 1998) (1998).
\bibitem{Suzumura_Berry}
Y.\ Suzumura and A.\  Kobayashi,
{\it private communicatiion}.
\bibitem{TakahashiBETS}
\SU{
K.\ Hiraki, S.\ Harada, K.\ Arai, Y.\ Takano, T.\ Takahashi:
 J.\ Phys.\ Soc.\ Jpn.\ \textbf{80}, 014715 (2011).
}
\bibitem{Kane-Mele2005}
C.L.\ Kane and  E.J.\ Mele,
 Phys.\ Rev.\ Lett.\  \textbf{95}, 226801 (2005).

\end{thebibliography}
\end{document}